\documentclass[preprint,12pt]{elsarticle}

\usepackage{amssymb}

\journal{Nuclear Physics B}

\begin{document}

\begin{frontmatter}

\title{Lepton-flavor violation and $(g-2)_\mu$ in the $\mu\nu$SSM}

\author[label1,label2]{Hai-Bin Zhang\corref{cor1}}
\ead{hbzhang@mail.dlut.edu.cn}
\cortext[cor1]{Corresponding author}
\author[label1,label2]{Tai-Fu Feng}
\ead{fengtf@hbu.edu.cn}
\author[label1]{Shu-Min Zhao}
\author[label1,label2]{Tie-Jun Gao}

\address[label1]{Department of Physics, Hebei University, Baoding 071002, China}
\address[label2]{Department of Physics, Dalian University of Technology, Dalian 116024, China }

\begin{abstract}
Within framework of the $\mu$ from $\nu$ Supersymmetric Standard Model ($\mu\nu$SSM), exotic singlet right-handed neutrino superfields induce new sources for lepton-flavor violation. In this work, we investigate some lepton-flavor violating processes in detail in the $\mu\nu$SSM. The numerical results indicate that the branching ratios for lepton-flavor violating processes $\mu\rightarrow e\gamma$, $\tau\rightarrow\mu\gamma$ and $\mu\rightarrow3e$ can reach $10^{-12}$ when $\tan\beta$ is large enough, which can be detected in near future. We also discuss the constraint on the relevant parameter space of the model from the muon anomalous magnetic dipole moment. In addition, from the scalars for the $\mu\nu$SSM we strictly separate the Goldstone bosons, which disappear in the physical gauge.
\end{abstract}


\end{frontmatter}

\section{Introduction}
\label{intro}
It is obviously evidence of new physics beyond the Standard Model (SM) that if we observe lepton-flavor violating (LFV) processes in future experiments, because the lepton-flavor number is conserved in the Standard Model. In supersymmetric (SUSY) extensions of the SM, the R-parity of a particle is defined as $R = (-1)^{L+3B+2S}$ \cite{R-parity} and can be violated if either the baryon number ($B$) or lepton number ($L$) is not conserved \cite{RPV,BRPV}, where $S$ denotes the spin of concerned component field. Note that $R=+1$ for particles and $-1$ for superparticles.

Differing from the models in Refs.\cite{RPV,BRPV}, the authors of Ref.\cite{mnSSM} propose a supersymmetric extension of the SM named as the ``$\mu$ from $\nu$ Supersymmetric Standard Model'' ($\mu\nu$SSM), which solves the $\mu$ problem \cite{m-problem} of the Minimal Supersymmetric Standard Model (MSSM) \cite{MSSM} through the lepton number and R-parity breaking couplings between the right-handed neutrino superfields and the Higgses  $\epsilon _{ab}{\lambda _i}\hat \nu _i^c\hat H_d^a\hat H_u^b$ in the superpotential. The effective $\mu$ term $\epsilon _{ab} \mu \hat H_d^a\hat H_u^b$ is generated spontaneously through right-handed sneutrino vacuum expectation values (VEVs), $\mu  = {\lambda _i}\left\langle {\tilde \nu _i^c} \right\rangle$, as the electroweak symmetry is broken (EWSB). Note that a popular model is the so-called Bilinear R-parity Violation (BRpV) model \cite{BRPV}, where the BRpV terms $\epsilon _{ab} \varepsilon_i \hat H_u^b\hat L_i^a$ are added to the MSSM. The effective  BRpV terms are generated spontaneously through the R-parity conserved terms $\epsilon _{ab}{Y_{{\nu _{ij}}}}\hat \nu _j^c \hat H_u^b\hat L_i^a$ in the superpotential of the $\mu\nu$SSM, and $\varepsilon_i= Y_{\nu _{ij}} \left\langle {\tilde \nu _j^c} \right\rangle$, as EWSB. So largely differing from the other models \cite{RPV,BRPV}, the $\mu\nu$SSM introduces three exotic right-handed sneutrinos $\hat{\nu}_i^c$, and once EWSB the right-handed sneutrinos give nonzero VEVs.  In addition, the nonzero VEVs of right-handed sneutrinos induce new sources for lepton-flavor violation. In this work, we analyze the constraints on parameter space of this model from the experimental observations on some LFV processes and muon anomalous magnetic dipole moment (MDM).

If the left-handed scalar neutrinos acquire nonzero vacuum expectation values when the electroweak symmetry is broken , the tiny neutrino masses are aroused \cite{Feng2} to account for the experimental data on neutrino oscillations \cite{oscillation1,oscillation2,oscillation3}.
Three flavor neutrinos $\nu_{e,\mu,\tau}$ are mixed into three massive neutrinos $\nu_{1,2,3}$ during their flight, and the mixings are described by the Pontecorvo-Maki-Nakagawa-Sakata unitary matrix
$U_{_{PMNS}}$ \cite{neutrino-oscillations}. The experimental observations of the parameters in $U_{_{PMNS}}$ for the normal mass hierarchy \cite{Valle1} show that \cite{PDG}
\begin{eqnarray}
&&\Delta m_{21}^2=7.58_{-0.26}^{+0.22}\times 10^{-5} {\rm eV}^2\;,\qquad\Delta m_{32}^2=2.35_{-0.09}^{+0.12}\times 10^{-3} {\rm eV}^2\;,\nonumber\\
&&\sin^2\theta_{12}=0.306_{-0.015}^{+0.018},\;\;\;\sin^2\theta_{23}=0.42_{-0.03}^{+0.08},\;\;\;
\sin^2\theta_{13}=0.021_{-0.008}^{+0.007}.
\label{neutrino-oscillations1}
\end{eqnarray}
Note that the Daya Bay Reactor Neutrino Experiment has measured a nonzero value for the neutrino mixing angle ${\theta _{13}}$ with a significance of 5.2 standard deviations recently \cite{Da-Ya-Bay}. Differing from the BRpV model, where one neutrino mass is generated at tree level and the other two at one loop \cite{BRpV-neutrino}, the $\mu\nu$SSM can generate three neutrino masses at the tree level through the mixing with the neutralinos including three right-handed neutrinos \cite{Roy,Bartl}.  Here,  we use the neutrino experimental data presented in Eq.(\ref{neutrino-oscillations1}) to restrain the input parameters in the model. Then, we analyze the branching ratios for the various LFV processes: $\mu\rightarrow e\gamma$, $\tau\rightarrow\mu\gamma$, $\mu\rightarrow3e$, etc., and the corrections to the anomalous magnetic dipole moment of the muon $a_\mu$ in the $\mu\nu$SSM.  The numerical results indicate that the new physics contributes large corrections to the branching ratios
of the mentioned LFV processes and $a_\mu$ in some parameter space of the model.

The outline of the paper is as follow. In section \ref{sec:2}, we present the ingredients of the $\mu\nu$SSM by introducing its superpotential and the general soft SUSY-breaking terms, in particular we strictly separate the unphysical Goldstone bosons from the scalars. In section \ref{sec:3}, we analyze the decay width of those interested rare LFV processes, and present the SUSY contribution to muon MDM in section \ref{sec:4}. The numerical analysis is given in section \ref{sec:5}, and the conclusions are summarized in section \ref{sec:6}. The tedious formulae are collected in Appendices.

\section{The $\mu\nu$SSM\label{sec:2}}
Besides the superfields of the MSSM, the $\mu\nu$SSM introduces three exotic gauge singlet neutrino
superfields $\hat{\nu}_i^c$. The corresponding superpotential of the $\mu\nu$SSM is given as \cite{mnSSM}
\begin{eqnarray}
&&W \:=\:{\epsilon _{ab}}\left( {{Y_{{u_{ij}}}}\hat H_u^b\hat Q_i^a\hat u_j^c + {Y_{{d_{ij}}}}\hat H_d^a\hat Q_i^b\hat d_j^c
+ {Y_{{e_{ij}}}}\hat H_d^a\hat L_i^b\hat e_j^c + {Y_{{\nu _{ij}}}}\hat H_u^b\hat L_i^a\hat \nu _j^c} \right)  \nonumber\\
&&\qquad\quad - \: {\epsilon _{ab}}{\lambda _i}\hat \nu _i^c\hat H_d^a\hat H_u^b + \frac{1}{3}{\kappa _{ijk}}\hat \nu _i^c\hat \nu _j^c\hat \nu _k^c \: ,
\end{eqnarray}
where $\hat H_d^T = \Big( {\hat H_d^0,\hat H_d^ - } \Big)$, $\hat H_u^T = \Big( {\hat H_u^ + ,\hat H_u^0} \Big)$, $\hat Q_i^T = \Big( {{{\hat u}_i},{{\hat d}_i}} \Big)$, $\hat L_i^T = \Big( {{{\hat \nu}_i},{{\hat e}_i}} \Big)$ (the index $T$ denotes the transposition) are $SU(2)$ doublet superfields, and $\hat d_j^c$, $\hat u_j^c$ and $\hat e_j^c$ represent the singlet down-type quark, up-type quark and lepton superfields, respectively.  In addition, $Y_{u,d,e,\nu}$, $\lambda$ and $\kappa$ are dimensionless matrices, a vector and a totally symmetric tensor.  $a,b$ are SU(2) indices with antisymmetric tensor $\epsilon_{12}=-\epsilon_{21}=1$, and $i,j=1,\;2,\;3$. The summation convention is implied on repeated indices.

In the superpotential, the first three terms are almost the same as the MSSM. Next two terms can generate the effective bilinear terms $\epsilon _{ab} \varepsilon_i \hat H_u^b\hat L_i^a$, $\epsilon _{ab} \mu \hat H_d^a\hat H_u^b$,  and $\varepsilon_i= Y_{\nu _{ij}} \left\langle {\tilde \nu _j^c} \right\rangle$, $\mu  = {\lambda _i}\left\langle {\tilde \nu _i^c} \right\rangle$,  once the electroweak symmetry is broken. The last term can generate the effective Majorana masses for neutrinos at the electroweak scale. And the last two terms explicitly violate lepton number and R-parity.

The general soft SUSY-breaking terms in the $\mu\nu$SSM are given by
\begin{eqnarray}
&&- \mathcal{L}_{soft}\:=\:m_{{{\tilde Q}_{ij}}}^{\rm{2}}\tilde Q{_i^{a\ast}}\tilde Q_j^a
+ m_{\tilde u_{ij}^c}^{\rm{2}}\tilde u{_i^{c\ast}}\tilde u_j^c + m_{\tilde d_{ij}^c}^2\tilde d{_i^{c\ast}}\tilde d_j^c
+ m_{{{\tilde L}_{ij}}}^2\tilde L_i^{a\ast}\tilde L_j^a \qquad\qquad\quad \nonumber\\
&&\quad + \: m_{\tilde e_{ij}^c}^2\tilde e{_i^{c\ast}}\tilde e_j^c + m_{{H_d}}^{\rm{2}} H_d^{a\ast} H_d^a
+ m_{{H_u}}^2H{_u^{a\ast}}H_u^a + m_{\tilde \nu_{ij}^c}^2\tilde \nu{_i^{c\ast}}\tilde \nu_j^c \nonumber\\
&&\quad + \: \epsilon_{ab}{\left[{{({A_u}{Y_u})}_{ij}}H_u^b\tilde Q_i^a\tilde u_j^c
+ {{({A_d}{Y_d})}_{ij}}H_d^a\tilde Q_i^b\tilde d_j^c + {{({A_e}{Y_e})}_{ij}}H_d^a\tilde L_i^b\tilde e_j^c + {\rm{H.c.}} \right]} \nonumber\\
&&\quad + \left[ {\epsilon _{ab}}{{({A_\nu}{Y_\nu})}_{ij}}H_u^b\tilde L_i^a\tilde \nu_j^c
- {\epsilon _{ab}}{{({A_\lambda }\lambda )}_i}\tilde \nu_i^c H_d^a H_u^b
+ \frac{1}{3}{{({A_\kappa }\kappa )}_{ijk}}\tilde \nu_i^c\tilde \nu_j^c\tilde \nu_k^c + {\rm{H.c.}} \right] \nonumber\\
&&\quad - \: \frac{1}{2}\left({M_3}{{\tilde \lambda }_3}{{\tilde \lambda }_3}
+ {M_2}{{\tilde \lambda }_2}{{\tilde \lambda }_2} + {M_1}{{\tilde \lambda }_1}{{\tilde \lambda }_1} + {\rm{H.c.}} \right).
\end{eqnarray}
Here, the front two lines contain squared-mass terms of squarks, sleptons and Higgses. The next two lines consist of the trilinear scalar couplings. In the last line, $M_3$, $M_2$ and $M_1$ denote Majorana masses corresponding to $SU(3)$, $SU(2)$ and $U(1)$ gauginos $\hat{\lambda}_3$, $\hat{\lambda}_2$ and $\hat{\lambda}_1$, respectively. In addition to the terms from $\mathcal{L}_{soft}$, the tree-level scalar potential receives the usual D and F term contributions \cite{mnSSM}.

When the electroweak symmetry is spontaneously broken (EWSB), the neutral scalars develop in general the vacuum expectation values (VEVs):
\begin{eqnarray}
\langle H_d^0 \rangle = \upsilon_d , \qquad \langle H_u^0 \rangle = \upsilon_u , \qquad
\langle \tilde \nu_i \rangle = \upsilon_{\nu_i} , \qquad \langle \tilde \nu_i^c \rangle = \upsilon_{\nu_i^c} .
\end{eqnarray}
Thus one can define neutral scalars as usual
\begin{eqnarray}
&&H_d^0=\frac{h_d + i P_d}{\sqrt{2}} + \upsilon_d\:, \qquad\;\quad H_u^0=\frac{h_u + i P_u}{\sqrt{2}} + \upsilon_u\:, \qquad \nonumber\\
&&\tilde \nu_i = \frac{(\tilde \nu_i)^R + i (\tilde \nu_i)^I}{\sqrt{2}} + \upsilon_{\nu_i}\:, \;\;\quad \tilde \nu_i^c = \frac{(\tilde \nu_i^c)^R + i (\tilde \nu_i^c)^I}{\sqrt{2}} + \upsilon_{\nu_i^c}\:.
\end{eqnarray}

For simplicity we will assume that all parameters in the potential are real in the following. After EWSB, the scalars mass matrices $M_S^2$, $M_P^2$ and $M_{S^{\pm}}^2$ are given in \ref{appendix-mass}. The CP-odd neutral scalars mass matrix $M_P^2$ contains a massless unphysical Goldstone boson $G^0$, which can be written as
\begin{eqnarray}
G^0 = {1 \over \sqrt{\upsilon_d^2+\upsilon_u^2+\upsilon_{\nu_i} \upsilon_{\nu_i}}} \Big(\upsilon_d {P_d}-\upsilon_u{P_u}+\upsilon_{\nu_i}{(\tilde \nu_i)^I}\Big)\:
\end{eqnarray}
with an $8\times8$ unitary matrix $Z_H$
\begin{eqnarray}
Z_H = \left( {\begin{array}{*{20}{c}}
   \frac{\upsilon_d}{\upsilon_{_{\rm{EW}}}} & \frac{\upsilon_u}{\upsilon_{_{\rm{SM}}}} & \frac{\upsilon_{\nu_1}\upsilon_d}{\upsilon_{_{\rm{EW}}}\upsilon_{_{\rm{SM}}}} & \frac{\upsilon_{\nu_2}\upsilon_d}{\upsilon_{_{\rm{EW}}}\upsilon_{_{\rm{SM}}}} & \frac{\upsilon_{\nu_3}\upsilon_d}{\upsilon_{_{\rm{EW}}}\upsilon_{_{\rm{SM}}}} & 0_{3\times1}   \\  [6pt]
   -\frac{\upsilon_u}{\upsilon_{_{\rm{EW}}}} & \frac{\upsilon_d}{\upsilon_{_{\rm{SM}}}} & -\frac{\upsilon_{\nu_1}\upsilon_u}{\upsilon_{_{\rm{EW}}}\upsilon_{_{\rm{SM}}}} & -\frac{\upsilon_{\nu_2}\upsilon_u}{\upsilon_{_{\rm{EW}}}\upsilon_{_{\rm{SM}}}} & -\frac{\upsilon_{\nu_3}\upsilon_u}{\upsilon_{_{\rm{EW}}}\upsilon_{_{\rm{SM}}}} & 0_{3\times1}   \\  [6pt]
   \frac{\upsilon_{\nu_1}}{\upsilon_{_{\rm{EW}}}} & 0 & -\frac{\upsilon_{_{\rm{SM}}}}{\upsilon_{_{\rm{EW}}}} & \frac{\upsilon_{\nu_3}}{\upsilon_{_{\rm{EW}}}} & -\frac{\upsilon_{\nu_2}}{\upsilon_{_{\rm{EW}}}} & 0_{3\times1} \\  [6pt]
   \frac{\upsilon_{\nu_2}}{\upsilon_{_{\rm{EW}}}} & 0 & -\frac{\upsilon_{\nu_3}}{\upsilon_{_{\rm{EW}}}} & -\frac{\upsilon_{_{\rm{SM}}}}{\upsilon_{_{\rm{EW}}}} & \frac{\upsilon_{\nu_1}}{\upsilon_{_{\rm{EW}}}} & 0_{3\times1}  \\  [6pt]
   \frac{\upsilon_{\nu_3}}{\upsilon_{_{\rm{EW}}}} & 0 & \frac{\upsilon_{\nu_2}}{\upsilon_{_{\rm{EW}}}} & -\frac{\upsilon_{\nu_1}}{\upsilon_{_{\rm{EW}}}} & -\frac{\upsilon_{_{\rm{SM}}}}{\upsilon_{_{\rm{EW}}}} & 0_{3\times1}  \\  [6pt]
   0_{1\times3} & 0_{1\times3} & 0_{1\times3} & 0_{1\times3} & 0_{1\times3} & 1_{3\times3}  \\  [6pt]
\end{array}} \right),
\end{eqnarray}
where $\upsilon_{_{\rm{SM}}}=\sqrt{\upsilon_d^2+\upsilon_u^2}$ and $\upsilon_{_{\rm{EW}}}=\sqrt{\upsilon_d^2+\upsilon_u^2+\upsilon_{\nu_i}\upsilon_{\nu_i}}$.   Making use of the minimization conditions of the tree-level neutral scalar potential, which are given in \ref{appendix-mini}, we have
\begin{eqnarray}
\left\{ {\begin{array}{l}
    {(Z_H^T M_P^2 Z_H)}_{11}=0\:,  \\ [6pt]
    {(Z_H^T M_P^2 Z_H)}_{1\alpha}={(Z_H^T M_P^2 Z_H)}_{\alpha1}=0,\quad \alpha=2,\ldots,8.  \\ [6pt]
\end{array}} \right.
\end{eqnarray}
The remaining $7\times7$ matrix $\Big({(Z_H^T M_P^2 Z_H)}_{\alpha\beta}\Big)\:(\alpha,\beta=2,\ldots,8)$ can be  further diagonalized, and then gives seven diagonal masses.
The charged scalars mass matrix $M_{S^{\pm}}^2$ also contains the massless unphysical Goldstone bosons $G^{\pm}$, which can be written as
\begin{eqnarray}
G^{\pm} = {1 \over \sqrt{\upsilon_d^2+\upsilon_u^2+\upsilon_{\nu_i} \upsilon_{\nu_i}}} \Big(\upsilon_d H_d^{\pm} - \upsilon_u {H_u^{\pm}}+\upsilon_{\nu_i}\tilde e_{L_i}^{\pm}\Big)\:
\end{eqnarray}
with the unitary matrix $Z_H$ and
\begin{eqnarray}
\left\{ {\begin{array}{l}
    {(Z_H^T M_{S^{\pm}}^2 Z_H)}_{11}=0\:,  \\ [6pt]
    {(Z_H^T M_{S^{\pm}}^2 Z_H)}_{1\alpha}={(Z_H^T M_{S^{\pm}}^2 Z_H)}_{\alpha1}=0,\quad \alpha=2,\ldots,8.  \\ [6pt]
\end{array}} \right.
\end{eqnarray}
In the physical (unitary) gauge, the Goldstone bosons $G^0$ and $G^{\pm}$ are eaten by $Z$-boson and $W$-boson, respectively, and disappear from the Lagrangian.

Then the mass squared of charged and neutral gauge boson are
\begin{eqnarray}
\left\{ {\begin{array}{l}
    m_W^2={e^2\over2s_{_W}^2}\Big(\upsilon_u^2+\upsilon_d^2+\upsilon_{\nu_i} \upsilon_{\nu_i}\Big),  \\ [6pt]
    m_Z^2={e^2\over {2s_{_W}^2 c_{_W}^2}}\Big(\upsilon_u^2+\upsilon_d^2+\upsilon_{\nu_i} \upsilon_{\nu_i}\Big),  \\ [6pt]
\end{array}} \right.
\end{eqnarray}
and
\begin{eqnarray}
\tan\beta={\upsilon_u\over\sqrt{\upsilon_d^2+\upsilon_{\nu_i}\upsilon_{\nu_i}}}\;.
\label{tanb}
\end{eqnarray}
Here $e$ is the electromagnetic coupling constant, $s_{_W}=\sin\theta_{_W}$ and $c_{_W}=\cos\theta_{_W}$ with $\theta_{_W}$ is the Weinberg angle, respectively.

\section{Lepton-flavor violation in the $\mu\nu$SSM\label{sec:3}}
In this section, we present the analysis on the decay width of the rare LFV processes $l_j^-\rightarrow l_i^-\gamma$ and $l_j^-  \rightarrow l_i^- l_i^- l_i^+$ in the $\mu\nu$SSM. For this study we will use the indices $\beta,\zeta=1,\ldots,5$, $\alpha,\rho=1,\ldots,8$, and $\eta,\sigma=1,\ldots,10$. And the summation convention is implied on the repeated indices.

\subsection{Rare decay $l_j^-\rightarrow l_i^-\gamma$}

\begin{figure}[htbp]
\setlength{\unitlength}{1mm}
\centering
\includegraphics[width=3.9in]{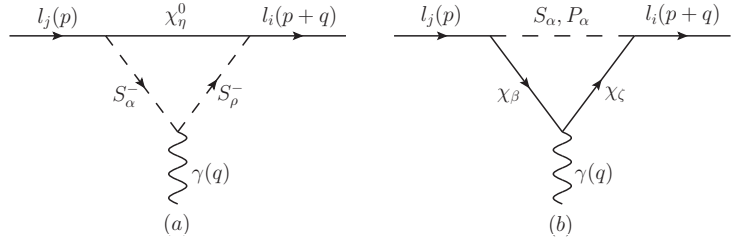}
\caption[]{Feynman diagrams for the LFV process $l_j^-\rightarrow l_i^-\gamma$. (a) represents the contributions from neutral fermions $\chi_\eta^0$ and charged scalars $S_{\alpha,\rho}^-$ loops, while (b) represents the contributions from charged fermions $\chi_{\beta,\zeta}$ and neutral scalars $N_\alpha$ ($N=S,P$) loops.}
\label{fig1}
\end{figure}

The amplitude for $l_j^-\rightarrow l_i^-\gamma$ (including $\mu\rightarrow e\gamma$
and $\tau\rightarrow\mu\gamma$ ) is generally written as \cite{Hisano}
\begin{eqnarray}
&&T = e{\epsilon ^\mu }{\bar u_i}(p + q)\Big[q^2{\gamma _\mu }(A_1^L{P_L} + A_1^R{P_R})
\nonumber\\
&&\qquad + \: {m_{{l_j}}}i{\sigma _{\mu \nu }}{q^\nu }(A_2^L{P_L} + A_2^R{P_R})\Big]{u_j}(p)\:,
\label{amplitude-gamma}
\end{eqnarray}
where $q$ is the injecting photon momentum, $p$ is the injecting lepton momentum, and $m_{{l_j}}$ is the mass of the $j$-th generation charged lepton, respectively. Furthermore, $\epsilon$ is the photon polarization vector, ${u_i}(p)$ (${v_i}(p)$ in the expressions below) is the wave function for lepton (antilepton), and $P_L=\frac{1}{2}{(1 - {\gamma _5})}$, $P_R=\frac{1}{2}{(1 + {\gamma _5})}$. Here, the Feynman diagrams contributing to the above amplitude are shown in Fig.\ref{fig1}. And the coefficients can be written by
\begin{eqnarray}
A_a^{L,R} = A_a^{(n)L,R} + A_a^{(c)L,R} \quad (a = 1,2)\:,
\end{eqnarray}
where $A_a^{(n)L,R}$ denote the contributions from the virtual neutral fermion loops, and $A_a^{(c)L,R}$ stand for the contributions from the virtual charged fermion loops, respectively. After integrating the heavy freedoms out, we formulate those coefficients as follows
\begin{eqnarray}
&&A_1^{(n)L} = \frac{1}{6{m_W^2}}C_R^{S_\alpha^- \chi_\eta^\circ {{\bar \chi }_i}}C_L^{S_\alpha^{-\ast} {\chi _j}\bar \chi _\eta ^ \circ }{I_4}({x_{\chi _\eta ^ \circ }},{x_{S_\alpha ^ -}})\;,\nonumber\\
&&A_2^{(n)L} = \frac{{{m_{\chi _\eta ^ \circ }}}}{{{m_{{l_j}}}}{m_W^2}}C_L^{S_\alpha ^ - \chi _\eta ^ \circ {{\bar \chi }_i}}C_L^{S_\alpha ^{-\ast} {\chi _j}\bar \chi _\eta ^ \circ }
\Big[ {I_3}({x_{\chi _\eta ^ \circ }},{x_{S_\alpha ^ - }}) - {I_1}(
{x_{\chi _\eta ^ \circ }},{x_{S_\alpha ^ - }}) \Big]
\;, \nonumber\\
&&A_a^{(n)R} = \left. {A_a^{(n)L}} \right|{ _{L \leftrightarrow R}}\:,
\end{eqnarray}
where the concrete expressions for form factors $I_k \:(k=1,\ldots,4)$ can be found in \ref{appendix-integral}.
Additionally, $x= {m^2}/{m_W^2}$, $m$ is the mass for the corresponding particle and $m_W$ is the mass for the $W$-boson, respectively. In a similar way, the corrections from the Feynman diagrams with virtual charged fermions are
\begin{eqnarray}
&&A_1^{(c)L} = \sum\limits_{N=S,P} \frac{1}{6{m_W^2}}
C_R^{{N_\alpha }{\chi _\beta }{{\bar \chi }_i}}C_L^{{N_\alpha }{\chi _j}{{\bar \chi }_\beta }}
\Big[ {I_1}({x_{{\chi _\beta }}},{x_{{N_\alpha }}})- 2 {I_2}({x_{{\chi _\beta }}},{x_{{N_\alpha }}})\nonumber\\
&&\qquad\qquad\qquad- {I_4}({x_{{\chi _\beta }}},{x_{{N_\alpha }}}) \Big] \;, \nonumber\\
&&A_2^{(c)L} = \sum\limits_{N=S,P} \frac{{{m_{{\chi _{^\beta }}}}}}{{{m_{{l_j}}}}{m_W^2}}
C_L^{{N_\alpha }{\chi _\beta }{{\bar \chi }_i}}
C_L^{{N_\alpha }{\chi _j}{{\bar \chi }_\beta }}\Big[ {I_1}(
{x_{{\chi _\beta }}},{x_{{N_\alpha }}}) - {I_2}({x_{{\chi _\beta }}},{x_{{N_\alpha }}}) \nonumber\\
&&\qquad\qquad\qquad - {I_4}({x_{{\chi _\beta }}},{x_{{N_\alpha }}}) \Big] \;, \nonumber\\
&&A_a^{(c)R} = \left. {A_a^{(c)L}} \right|{ _{L \leftrightarrow R}}\:.
\end{eqnarray}

Using the amplitude presented in Eq.(\ref{amplitude-gamma}), we then obtain the decay width for $l_j^-\rightarrow l_i^-\gamma$ as \cite{Hisano}
\begin{eqnarray}
\Gamma (l_j^ -  \to l_i^ - \gamma ) = \frac{{{e^2}}}{{16\pi }}m_{{l_j}}^5 \Big({\left| {A_2^L} \right|^2} + {\left| {A_2^R} \right|^2}\Big)\:.
\label{gamma-1}
\end{eqnarray}
And the branching ratio of $l_j^-\rightarrow l_i^-\gamma$ is
\begin{eqnarray}
{\rm{Br}}(l_j^-  \to l_i^ - \gamma ) = \frac{\Gamma (l_j^ -  \to l_i^ - \gamma )}{\Gamma_{l_j^-}}\:,
\end{eqnarray}
where $\Gamma_{l_j^-}$ denotes the total decay rate of the lepton $l_j^-$. In the numerical calculation, $\Gamma_\mu \approx 2.996 \times 10^{-19}\:{\rm{GeV}}$ for the muon and $\Gamma_\tau \approx 2.265 \times 10^{-12}\:{\rm{GeV}}$ for the tauon.

\subsection{Rare decay $l_j^- \rightarrow l_i^- l_i^- l_i^+$}
\indent\indent

\begin{figure}[htbp]
\setlength{\unitlength}{1mm}
\centering
\includegraphics[width=2.0in]{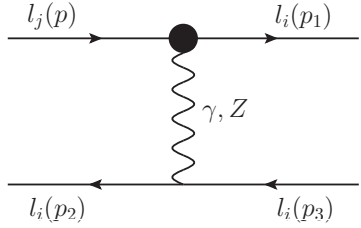}
\caption[]{Penguin-type diagrams for the LFV process $l_j^- \rightarrow l_i^- l_i^- l_i^+$ in which a photon $\gamma$ and $Z$-boson are exchanged. The blob indicates an $l_j^- - l_i^- - \gamma$ vertex such as Fig.\ref{fig1} or $l_j^- - l_i^- - Z$ vertex where the $Z$-boson is external.}
\label{fig2}
\end{figure}
For the rare LFV processes $l_j^-  \rightarrow l_i^- l_i^- l_i^+$ (including $\mu\rightarrow3e$), the corresponding effective Hamilton originates from penguin-type diagrams and from box-type diagrams. The $\gamma$-penguin contribution can be computed using Eq.(\ref{amplitude-gamma}), with the result
\begin{eqnarray}
&&T_{\gamma - {\rm{p}}} = {\bar u_i}({p_1})\Big[{q^2}{\gamma _\mu }(A_1^L{P_L}
+ A_1^R{P_R}) + {m_{l_j}}i{\sigma _{\mu \nu }}{q^\nu }(A_2^L{P_L}
+ A_2^R{P_R})\Big] {u_j}(p)  \nonumber\\
&&\qquad\quad\;\; \times \: \frac{{{e^2}}}{{{q^2}}}{\bar u_i}({p_2}){\gamma ^\mu }
{v_i}({p_3})- ({p_1} \leftrightarrow {p_2})\:.
\end{eqnarray}
Similarly, the contribution from $Z$-penguin diagrams which are depicted by Fig.\ref{fig2} is
\begin{eqnarray}
&&T_{Z- {\rm{p}}} = \frac{{{e^2}}}{{m_Z^2}}{\bar u_i}({p_1}){\gamma _\mu }({F_L}{P_L} + {F_R}{P_R}){u_j}(p)  {\bar u_i}({p_2}){\gamma ^\mu }\Big(C_L^{Z{\chi _{2 + i}}{\bar{\chi} _{2 + i}}}{P_L} \qquad \nonumber\\
&&\qquad\quad\;\; + \: C_R^{Z {\chi _{2 + i}}{\bar{\chi} _{2 + i}}}{P_R}\Big){v_i}({p_3}) - ({p_1} \leftrightarrow {p_2})\:,
\end{eqnarray}
where $m_Z$ is the mass for the $Z$-boson and
\begin{eqnarray}
{F_{L,R}} = F_{L,R}^{(n)} + F_{L,R}^{(c)}\:.
\end{eqnarray}
The contributions to the effective couplings $F_{L,R}^{(n)}$ and $F_{L,R}^{(c)}$ are
\begin{eqnarray}
&&F_L^{(n)} = \sum\limits_{N=S,P} \Big[ \frac{{m_{{\chi _\zeta }}}{m_{{\chi _\beta }}}}
{{e^2}{m_W^2}}C_R^{{N_\alpha }{\chi _\zeta }{{\bar \chi }_i}}C_L^{Z{\chi _\beta }{{\bar \chi }_\zeta }}
C_L^{{N_\alpha }{\chi _j}{{\bar \chi }_\beta }}{G_1}({x_{{N_\alpha }}},{x_{{\chi _\zeta }}},{x_{{\chi _\beta }}})\nonumber\\
&&\qquad\;\quad - \: \frac{1}{2{e^2}} C_R^{{N_\alpha }{\chi _\zeta }{{\bar \chi }_i}}C_R^{Z{\chi _\beta }
{{\bar \chi }_\zeta }}C_L^{{N_\alpha }{\chi _j}{{\bar \chi }_\beta }}{G_2}({x_{{N_\alpha }}},
{x_{{\chi _\zeta }}},{x_{{\chi _\beta }}}) \Big]\:,
\nonumber\\
&&F_L^{(c)} = \, \frac{1}{2{e^2}}C_R^{S_\rho ^ - \chi _\eta ^0{{\bar \chi }_i}}
C_R^{ZS_\alpha ^ - S_\rho ^ {-\ast} }C_L^{S_\alpha ^{-\ast} {\chi _j}\bar \chi _\eta ^0}{G_2}
({x_{\chi _\eta ^0}},{x_{S_\alpha ^ - }},{x_{S_\rho ^ - }}) \:,
\nonumber\\
&&F_R^{(n,c)} = \left. {F_L^{(n,c)}} \right|{ _{L \leftrightarrow R}} \:.
\end{eqnarray}
Here, the concrete expressions for $G_k$ are given in \ref{appendix-integral}.

\begin{figure}[htbp]
\setlength{\unitlength}{1mm}
\centering
\begin{minipage}[c]{0.79\textwidth}
\includegraphics[width=4.2in]{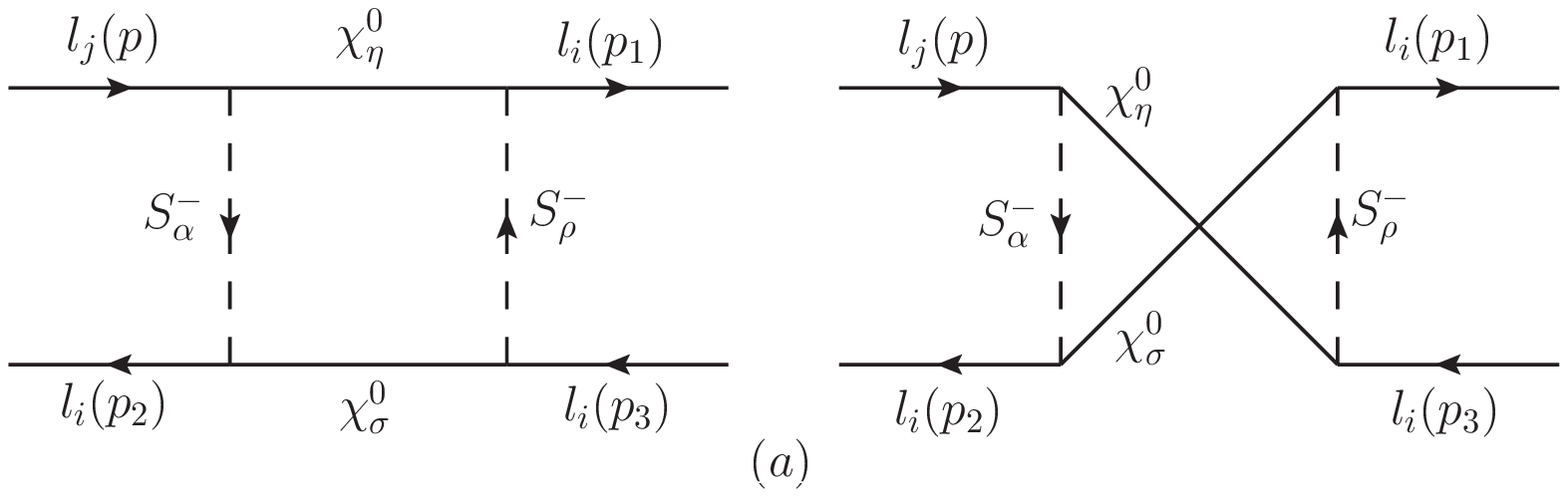}
\label{fig3a}
\end{minipage}
\begin{minipage}[c]{0.4\textwidth}
\includegraphics[width=2.1in]{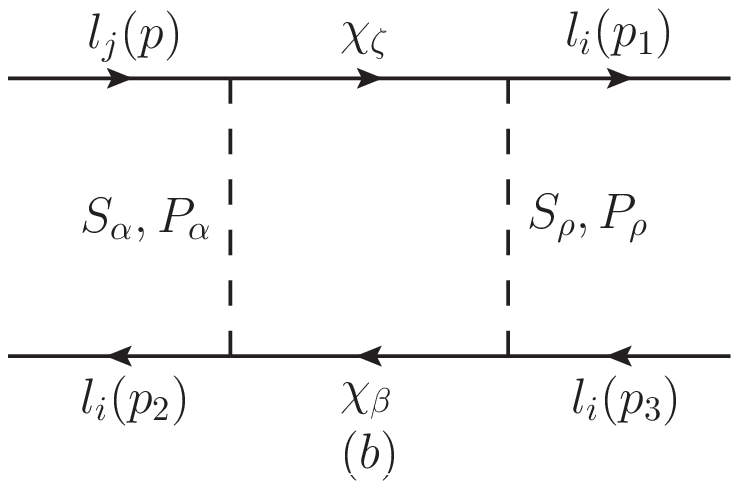}
\label{fig3b}
\end{minipage}
\caption[]{Box-type diagrams for the LFV process $l_j^- \rightarrow l_i^- l_i^- l_i^+$. (a) represents the contributions from neutral fermions $\chi_{\eta,\sigma}^0$ and charged scalars $S_{\alpha,\rho}^-$ loops, and (b) represents the contributions from charged fermions $\chi_{\beta,\zeta}$ and neutral scalars $N_{\alpha,\rho}$ ($N=S,P$) loops.}
\label{fig3}
\end{figure}

Furthermore, the effective Hamilton from the box-type diagrams which are drawn in Fig.\ref{fig3} can be written as
\begin{eqnarray}
&&T_{box} = \Big\{B_1^L{e^2}{\bar u_i}({p_1}){\gamma _\mu }{P_L}{u_j}(p){\bar u_i}(p_2)
{\gamma ^\mu }{P_L}{v_i}({p_3}) + (L \leftrightarrow R)\Big\}   \nonumber\\
&&\quad + \: \Big\{B_2^L{e^2}\Big[{\bar u_i}(p_1){\gamma _\mu }{P_L}{u_j}(p){\bar u_i}(p_2)
{\gamma ^\mu }{P_R}{v_i}({p_3}) - (p_1 \leftrightarrow {p_2})\Big] + {(L \leftrightarrow R) } \Big\} \nonumber\\
&&\quad + \: \Big\{ B_3^L{e^2}\Big[{\bar u_i}({p_1}){P_L}{u_j}(p){\bar u_i}({p_2}){P_L}{v_i}({p_3}) - ({p_1} \leftrightarrow {p_2})\Big] + (L \leftrightarrow R) \Big\}   \nonumber\\
&&\quad + \: \Big\{ B_4^L{e^2}\Big[{\bar u_i}({p_1}){\sigma _{\mu \nu }}{P_L}{u_j}(p){\bar u_i}({p_2}){\sigma ^{\mu \nu }}{P_L}{v_i}({p_3})
- ({p_1} \leftrightarrow {p_2})\Big]  + \: (L \leftrightarrow R)\Big\} \:\nonumber\\
&&
\end{eqnarray}
with
\begin{eqnarray}
B_a^{L,R} = B_a^{(n)L,R} + B_a^{(c)L,R} \quad (a = 1, \ldots ,4)\:.
\end{eqnarray}
The effective couplings $B_a^{(n)L,R}$ originate from those box diagrams with virtual neutral fermion contributions:
\begin{eqnarray}
&&B_1^{(n)L} = \frac{{m_{\chi_\eta^0}}{m_{\chi_\sigma^0}}}{{e^2}{m_W^4}}
{G_3}({x_{\chi_\eta^0}},{x_{\chi_\sigma^0}},{x_{S_\alpha^-}},
{x_{S_\rho^-}})C_L^{S_\rho ^ - \chi _\eta^0{{\bar \chi }_i}}
C_L^{S_\alpha^{-\ast} {\chi _j}\bar \chi _\eta^0}C_R^{S_\rho ^ - \chi _\sigma^0{{\bar \chi }_i}}
C_R^{S_\alpha^{-\ast} {\chi _i}\bar \chi _\sigma^0}
\nonumber\\
&&\qquad\;\quad +\:\frac{1}{2{e^2}{m_W^2}}{G_4}({x_{\chi_\eta^0}},{x_{\chi_\sigma^0}},{x_{S_\alpha^-}},
{x_{S_\rho^-}})\Big[
C_R^{S_\rho ^ - \chi _\eta^0{{\bar \chi }_i}}C_L^{S_\alpha^{-\ast} {\chi _j}\bar \chi _\eta^0}
C_R^{S_\alpha^ - \chi _\sigma^0{{\bar \chi }_i}}C_L^{S_\rho ^{-\ast} {\chi _i}\bar \chi _\sigma^0}
\nonumber\\
&&\qquad\;\quad + \:C_L^{S_\rho ^ - \chi _\eta^0{{\bar \chi }_i}}
C_R^{S_\alpha^{-\ast} {\chi _j}\bar \chi _\eta^0}C_R^{S_\rho ^ - \chi _\sigma^0{{\bar \chi }_i}}
C_L^{S_\alpha^{-\ast} {\chi _i}\bar \chi _\sigma^0}\Big]  ,
\nonumber\\
&&B_2^{(n)L} = - \frac{{m_{\chi_\eta^0}}{m_{\chi_\sigma^0}}}
{2{e^2}{m_W^4}}{G_3}({x_{\chi_\eta^0}},{x_{\chi_\sigma^0}},{x_{S_\alpha^-}},
{x_{S_\rho^-}})C_R^{S_\rho ^ - \chi _\eta^0{{\bar \chi }_i}}
C_R^{S_\alpha^{-\ast} {\chi _j}\bar \chi _\eta^0}C_L^{S_\alpha^ - \chi _\sigma^0{{\bar \chi }_i}}
C_L^{S_\rho ^{-\ast} {\chi _i}\bar \chi _\sigma^0}
\nonumber\\
&&\qquad\;\quad +\:\frac{1}{4{e^2}{m_W^2}}{G_4}({x_{\chi_\eta^0}},{x_{\chi_\sigma^0}},{x_{S_\alpha^-}},
{x_{S_\rho^-}})\Big[ C_R^{S_\rho ^ - \chi _\eta^0{{\bar \chi }_i}}
C_L^{S_\alpha^{-\ast} {\chi _j}\bar \chi _\eta^0}C_L^{S_\alpha^ - \chi _\sigma^0{{\bar \chi }_i}}
C_R^{S_\rho ^{-\ast} {\chi _i}\bar \chi _\sigma^0}
\nonumber\\
&&\qquad\;\quad + \:C_R^{S_\rho ^ - \chi _\eta^0{{\bar \chi }_i}}
C_L^{S_\alpha^{-\ast} {\chi _j}\bar \chi _\eta^0}C_R^{S_\rho ^ - \chi _\sigma^0{{\bar \chi }_i}}
C_L^{S_\alpha^{-\ast} {\chi _i}\bar \chi _\sigma^0}\Big]   ,
\nonumber\\
&&B_3^{(n)L}=\frac{{m_{\chi_\eta^0}}{m_{\chi_\sigma^0}}}{{e^2}{m_W^4}}{G_3}({x_{\chi_\eta^0}},{x_{\chi_\sigma^0}},{x_{S_\alpha^-}},
{x_{S_\rho^-}})\Big[
C_L^{S_\rho ^ - \chi _\eta^0{{\bar \chi }_i}}C_L^{S_\alpha^{-\ast} {\chi _j}\bar
\chi _\eta^0}C_L^{S_\alpha^ - \chi _\sigma^0{{\bar \chi }_i}}
C_L^{S_\rho ^{-\ast} {\chi _i}\bar \chi _\sigma^0}
\nonumber\\
&&\qquad\;\quad - \:\frac{1}{2}\:C_L^{S_\rho ^ - \chi _\eta^0{{\bar \chi }_i}}
C_L^{S_\alpha^{-\ast} {\chi _j}\bar \chi _\eta^0}C_L^{S_\rho ^ - \chi _\sigma^0{{\bar \chi }_i}}
C_L^{S_\alpha^{-\ast} {\chi _i}\bar \chi _\sigma^0}\Big] ,
\nonumber\\
&&B_4^{(n)L}=\frac{{m_{\chi_\eta^0}}{m_{\chi_\sigma^0}}}{8{e^2}{m_W^4}}{G_3}({x_{\chi_\eta^0}},{x_{\chi_\sigma^0}},{x_{S_\alpha^-}},
{x_{S_\rho^-}})
C_L^{S_\rho ^ - \chi _\eta^0{{\bar \chi }_i}}C_L^{S_\alpha^{-\ast} {\chi _j}\bar \chi _\eta^0}
C_L^{S_\rho ^ - \chi _\sigma^0{{\bar \chi }_i}}C_L^{S_\alpha^{-\ast} {\chi _i}\bar \chi _\sigma^0} ,
\nonumber\\
&&B_a^{(n)R} = \left. {B_a^{(n)L}} \right|
{ _{L \leftrightarrow R}} \:.
\end{eqnarray}
Correspondingly, the effective couplings from the box diagrams with virtual charged fermion
contributions $B_a^{(c)L,R}$ are
\begin{eqnarray}
&&B_1^{(c)L} = \sum\limits_{N=S,P} \frac{1}{2{e^2}{m_W^2}}
{G_4}({x_{{\chi _\zeta}}},{x_{{\chi _\beta }}},{x_{{N_\alpha}}},{x_{{N_\rho }}})C_R^{{N_\rho }{\chi _\zeta}{{\bar \chi }_i}}
C_L^{{N_\alpha}{\chi _j}{{\bar \chi }_\zeta}}C_R^{{N_\alpha}{\chi _\beta }{{\bar \chi }_i}}
C_L^{{N_\rho }{\chi _i}{{\bar \chi }_\beta }},\nonumber\\
&&B_2^{(c)L} = \sum\limits_{N=S,P} \Big[ \frac{1}{4{e^2}{m_W^2}} {G_4}({x_{{\chi _\zeta}}},{x_{{\chi _\beta }}}
,{x_{{N_\alpha}}},{x_{{N_\rho }}})C_R^{{N_\rho }{\chi _\zeta}{{\bar \chi }_i}}
C_L^{{N_\alpha}{\chi _j}{{\bar \chi }_\zeta}}C_L^{{N_\alpha}{\chi _\beta }{{\bar \chi }_i}}
C_R^{{N_\rho }{\chi _i}{{\bar \chi }_\beta }}\nonumber\\
&&\qquad\;\;\quad - \frac{{m_{\chi _\zeta}}{m_{\chi _\beta}}}{2{e^2}{m_W^4}}
{G_3}({x_{{\chi _\zeta}}},{x_{{\chi _\beta }}},{x_{{N_\alpha}}},
{x_{{N_\rho }}})C_R^{{N_\rho }{\chi _\zeta}{{\bar \chi }_i}}
C_R^{{N_\alpha}{\chi _j}{{\bar \chi }_\zeta}}C_L^{{N_\alpha}{\chi _\beta }{{\bar \chi }_i}}
C_L^{{N_\rho }{\chi _i}{{\bar \chi }_\beta }} \Big] ,
\nonumber\\
&&B_3^{(c)L} = \sum\limits_{N=S,P} \frac{{m_{\chi _\zeta}}
{m_{\chi _\beta}}}{{e^2}{m_W^4}} {{G_3}({x_{{\chi _\zeta}}}
,{x_{{\chi _\beta }}},{x_{{N_\alpha}}},{x_{{N_\rho }}})
C_L^{{N_\rho }{\chi _\zeta}{{\bar \chi }_i}}C_L^{{N_\alpha}{\chi _j}{{\bar \chi }_\zeta}}
C_L^{{N_\alpha}{\chi _\beta }{{\bar \chi }_i}}C_L^{{N_\rho }{\chi _i}{{\bar \chi }_\beta }}},
\nonumber\\
&&B_4^{(c)L} = 0\:,   \nonumber\\
&&B_a^{(c)R} = \left. {B_a^{(c)L}} \right|{ _{L \leftrightarrow R}}\:.
\end{eqnarray}

Using the expression for the above amplitude, we can calculate the decay width for $l_j^-  \rightarrow l_i^- l_i^- l_i^+$ \cite{Hisano}:
\begin{eqnarray}
&&\Gamma (l_j^ -  \to l_i^ - l_i^ - l_i^+) = \frac{{{e^4}}}{{512{\pi ^3}}}m_{{l_j}}^5  \Big\{
({\left| {A_2^L} \right|^2} + {\left| {A_2^R} \right|^2})(\frac{{16}}{3}\ln \frac{{{m_{{l_j}}}}}{{2{m_{{l_i}}}}} - \frac{{14}}{9}) \nonumber\\
&&\quad + \: ({\left| {A_1^L} \right|^2} + {\left| {A_1^R} \right|^2}) - 2(A_1^LA_2^{R * } + A_2^LA_1^{R * } + \textrm{H.c.}) + \frac{1}{6}({\left| {B_1^L} \right|^2}  + {\left| {B_1^R} \right|^2}) \nonumber\\
&&\quad  + \: \frac{1}{3}({\left| {B_2^L} \right|^2} + {\left| {B_2^R} \right|^2}) + \frac{1}{{24}}({\left| {B_3^L} \right|^2} + {\left| {B_3^R} \right|^2}) + 6({\left| {B_4^L} \right|^2} + {\left| {B_4^R} \right|^2})   \nonumber\\
&&\quad  - \: \frac{1}{2}(B_3^LB_4^{L * } + B_3^RA_4^{R * } + \textrm{H.c.}) + \frac{1}{3}(A_1^LB_1^{L * } + A_1^RB_1^{R * } + A_1^LB_2^{L * }  \nonumber\\
&&\quad  + \: A_1^RB_2^{R * } + \textrm{H.c.}) - \frac{2}{3}(A_2^RB_1^{L * } + A_2^LB_1^{R * }  + A_2^LB_2^{R * } + A_2^RB_2^{L * } + \textrm{H.c.})  \nonumber\\
&&\quad + \: \frac{1}{3}\Big[2({\left| {{F_{LL}}} \right|^2} + {\left| {{F_{RR}}} \right|^2}) + ({\left| {{F_{LR}}} \right|^2} + {\left| {{F_{RL}}} \right|^2}) + (B_1^LF_{LL}^ *  + B_1^RF_{RR}^ *   \nonumber\\
&&\quad  + \: B_2^LF_{LR}^ *  + B_2^RF_{RL}^ *  + \textrm{H.c.}) + 2(A_1^LF_{LL}^ *  + A_1^RF_{RR}^ *  + \textrm{H.c.})  \nonumber\\
&&\quad + \: (A_1^LF_{LR}^ *  + A_1^RF_{RL}^ *  + \textrm{H.c.}) - 4(A_2^RF_{LL}^ *  + A_2^LF_{RR}^ *  + \textrm{H.c.})  \nonumber\\
&&\quad  - \: 2(A_2^LF_{RL}^ *  + A_2^RF_{LR}^ *  + \textrm{H.c.})\Big]\Big\}\;
\label{gamma-2}
\end{eqnarray}
with
\begin{eqnarray}
&&{F_{LL}} = \frac{{{F_L}C_L^{Z{\chi _{2 + i}}{\bar{\chi} _{2 + i}}}}}{{m_Z^2}},\qquad
{F_{RR}} = {F_{LL}}\left| {_{L \leftrightarrow R}} \right., \quad \nonumber\\
&&{F_{LR}} = \frac{{{F_L}C_R^{Z{\chi _{2 + i}}{\bar{\chi} _{2 + i}}}}}{{m_Z^2}},\qquad
{F_{RL}} = {F_{LR}}\left| {_{L \leftrightarrow R}} \right.. \quad
\end{eqnarray}
And the branching ratio of $l_j^ -  \to l_i^ - l_i^ - l_i^+$ is
\begin{eqnarray}
{\rm{Br}}(l_j^ -  \to l_i^ - l_i^ - l_i^+) = \frac{\Gamma (l_j^ -  \to l_i^ - l_i^ - l_i^+)}{\Gamma_{l_j^-}}\:.
\end{eqnarray}

\section{$(g-2)_\mu$ in the $\mu\nu$SSM\label{sec:4}}
The anomalous magnetic dipole moment (MDM) of the muon can be actually be written as the operator
\begin{eqnarray}
\mathcal{L}_{MDM}=\frac{e}{4 m_\mu} a_\mu \bar{l}_\mu \sigma^{\alpha\beta} l_\mu F_{\alpha\beta}\:,
\end{eqnarray}
where $\sigma^{\alpha\beta}=\frac{i}{2}[\gamma^\alpha,\gamma^\beta]$, $F_{\alpha\beta}$ is the electromagnetic field strength, $l_\mu$ denotes the muon which is on-shell, $m_\mu$ is the muon mass and $a_\mu=\frac{1}{2}(g-2)_\mu$. Adopting the effective Lagrangian approach, we can get \cite{Feng3}
\begin{eqnarray}
a_\mu = \frac{4 Q_f m_{\mu}^2}{{(4 \pi)}^2} \Re{(C_2^R + C_2^{L\ast} + C_6^R)}\:,
\end{eqnarray}
where $Q_f=-1$, $\Re(\cdots)$ represents the operation to take the real part of a complex number
and $C_{2,6}^{L,R}$ denote the Wilson coefficients of the corresponding operators $O_{2,6}^{L,R}$
\begin{eqnarray}
&&O_2^{L,R} = \frac{e Q_f}{{(4 \pi)}^2} \overline{(i \mathcal{D}_\alpha l_\mu )}
\gamma^\alpha F\cdot \sigma P_{L,R} l_\mu\:, \nonumber\\
&&O_6^{L,R} = \frac{e Q_f m_{\mu}}{{(4 \pi)}^2} \overline{l}_\mu F\cdot \sigma P_{L,R} l_\mu\:.
\end{eqnarray}

In the $\mu\nu$SSM, the SUSY corrections can be written as
\begin{eqnarray}
C_{2,6}^{L,R}=C_{2,6}^{{L,R}(n)}+C_{2,6}^{{L,R}(c)}\:.
\end{eqnarray}
The effective couplings $C_{2,6}^{{L,R}(n)}$ represent the contributions from the triangle diagrams with virtual neutralinos
\begin{eqnarray}
&&C_2^{R(n)}=\frac{{(4 \pi)}^2}{Q_f {m_W^2}}C_L^{S_\alpha ^ - \chi _\eta ^ \circ {{\bar \chi }_4}}
C_R^{S_\alpha ^{-\ast} {\chi _4}\bar \chi _\eta ^ \circ }\Big[  - {I_3}{\rm{(}}{x_{\chi _\eta ^ \circ }},{x_{S_\alpha ^ - }}) + {I_4}({x_{\chi _\eta ^ \circ }},{x_{S_\alpha ^ - }}) \Big]\:,
\nonumber\\
&&C_6^{R(n)}=\frac{{(4 \pi)}^2 {m_{\chi _\eta ^ \circ }}}{Q_f {m_W^2}{m_\mu }}
C_R^{S_\alpha ^ - \chi _\eta ^ \circ {{\bar \chi }_4}}C_R^{S_\alpha ^{-\ast} {\chi _4}\bar \chi _\eta ^ \circ }
\Big[ - 2 {I_1}({x_{\chi _\eta ^ \circ }},{x_{S_\alpha ^ - }}) + 2 {I_3}({x_{\chi _\eta ^ \circ }},{x_{S_\alpha ^ - }}) \Big]\:,  \nonumber\\
&&C_{2,6}^{L(n)}=C_{2,6}^{R(n)}\mid _{L \leftrightarrow R}\:.
\end{eqnarray}
Similarly, the contributions $C_{2,6}^{{L,R}(c)}$ originating from triangle diagrams with virtual charginos are
\begin{eqnarray}
&&C_2^{R(c)}= \sum\limits_{N=S,P} \frac{{(4 \pi)}^2}{Q_f {m_W^2}} C_R^{{N_\alpha }{\chi _\beta }{{\bar \chi }_4}}C_L^{{N_\alpha }{\chi _4}{{\bar \chi }_\beta }}\Big[  - {I_1}({x_{{\chi _\beta }}},{x_{{N_\alpha }}}) + 2{I_3}({x_{{\chi _\beta }}},{x_{{N_\alpha }}})
\nonumber\\
&&\qquad\qquad\qquad - {I_4}({x_{{\chi _\beta }}},{x_{{N_\alpha }}}) \Big] \:,
\nonumber\\
&&C_6^{R(c)}= \sum\limits_{N=S,P} \frac{{(4 \pi)}^2 {m_{{\chi _{^\beta }}}}}{Q_f {m_W^2}{m_\mu }}
C_R^{{N_\alpha }{\chi _\beta }{{\bar \chi }_4}}C_R^{{N_\alpha }{\chi _4}{{\bar \chi }_\beta }}\Big[ 2 {I_1}({x_{{\chi _\beta }}},{x_{{N_\alpha }}}) - 2 {I_2}({x_{{\chi _\beta }}},{x_{{N_\alpha }}}) \nonumber\\
&&\qquad\qquad\qquad - 2 {I_3}({x_{{\chi _\beta }}},{x_{{N_\alpha }}}) \Big] \:, \nonumber\\
&&C_{2,6}^{L(c)}=C_{2,6}^{R(c)}\mid _{L \leftrightarrow R}\,.
\end{eqnarray}

\section{The numerical results\label{sec:5}}

\subsection{The parameter space}
It is well known that there are many free parameters in various SUSY extensions of the SM.
In order to obtain a more transparent numerical results, we take some assumptions on parameter
space of the $\mu\nu{\rm SSM}$ before we perform the numerical analysis.

In lepton sector, we adopt the minimal flavor violation (MFV) assumptions
\begin{eqnarray}
&&{\kappa _{ijk}} = \kappa  \textrm{ and }  {({A_\kappa }\kappa )_{ijk}} =
{A_\kappa }\kappa,\, \textrm{if } i=j=k,\, \textrm{and zero otherwise},  \nonumber\\
&&m_{{{\tilde L}_{ij}}}^2 = m_{\tilde L_i}^2{\delta _{ij}},\:
m_{\tilde \nu_{ij}^c}^2 = m_{{{\tilde \nu_i}^c}}^2{\delta _{ij}},\:
m_{\tilde e_{ij}^c}^2 = m_{{{\tilde e}^c}}^2{\delta _{ij}},  \nonumber\\
&&{Y_{{\nu _{ij}}}} = {Y_{{\nu _i}}}{\delta _{ij}}, \: {Y_{{e_{ij}}}} = {Y_{{e_i}}}{\delta _{ij}},
 \: \lambda _i = \lambda ,\: \upsilon_{\nu_i^c}=\upsilon_{\nu^c}, \nonumber\\
&&(A_\nu Y_\nu)_{ij}={A_\nu}{Y_{{\nu_i}}}{\delta _{ij}},\: {({A_e}{Y_e})_{ij}} = {A_e}{Y_{{e_i}}}{\delta _{ij}}\:,\:\textrm{and }{{\rm{(}}{A_\lambda }
\lambda {\rm{)}}_i} = {A_\lambda }\lambda,
\label{MFV}
\end{eqnarray}
where $i,\;j,\;k =1,\;2,\;3 $.

The $3\times3$ matrix $Y_\nu$ determines the Dirac masses for the neutrinos ${Y_\nu}{\upsilon_u} \sim {m_D}$, and the tiny neutrino masses are obtained through TeV scale seesaw mechanism $m_\nu\sim m_D m_{_N}^{-1}m_D^T$. This indicates that the nonzero VEVs of left-handed sneutrinos satisfy $\upsilon_{\nu_i}\ll\upsilon_{u,d}$, then
\begin{eqnarray}
\tan\beta\simeq \frac{\upsilon_u}{\upsilon_d}\:.
\end{eqnarray}

Assuming that the charged lepton mass matrix in the flavor basic is in the diagonal form, we get
\begin{eqnarray}
{Y_{{e_i}}} = \frac{{{m_{{l_i}}}}}{{{\upsilon_d}}},
\end{eqnarray}
where $m_{l_i}$ is the charged lepton $l_i$ mass, and we parameterize the unitary matrix which diagonalizes the effective light neutrino mass matrix $m_{eff}$ (can be found in \ref{appendix-approximate}) as \cite{Bilenky}
\begin{eqnarray}
{U_\nu} = &&\left( {\begin{array}{*{20}{c}}
   {{c_{12}}{c_{13}}} & {{s_{12}}{c_{13}}} & {{s_{13}}{e^{ - i\delta }}}  \\
   { - {s_{12}}{c_{23}} - {c_{12}}{s_{23}}{s_{13}}{e^{i\delta }}} & {{c_{12}}{c_{23}} - {s_{12}}
   {s_{23}}{s_{13}}{e^{i\delta }}} & {{s_{23}}{c_{13}}}  \\
   {{s_{12}}{s_{23}} - {c_{12}}{c_{23}}{s_{13}}{e^{i\delta }}} & { - {c_{12}}{s_{23}} - {s_{12}}
   {c_{23}}{s_{13}}{e^{i\delta }}} & {{c_{23}}{c_{13}}}  \\
\end{array}} \right)\nonumber\\
&&\: \times \: diag(1,{e^{i\frac{{{\alpha _{21}}}}{2}}},{e^{i\frac{{{\alpha _{31}}}}{2}}})\:,
\label{PMNS-matrix}
\end{eqnarray}
where ${c_{ij}} = \cos {\theta _{ij}}$, ${s_{ij}} = \sin {\theta _{ij}}$, the angles
${\theta _{ij}} = \left[\: {0,\pi/2} \:\right]$, $\delta = \left[ \:{0,2\pi } \:\right]$
is the Dirac CP violation phase and $\alpha_{21}$, $\alpha_{31}$ are two Majorana CP violation phases. Here,  we choose $\delta=\alpha_{21}=\alpha_{31}=0$.
$U_\nu$ diagonalizes $m_{eff}$ in the following way:
\begin{eqnarray}
U_\nu ^T m_{eff}^T{m_{eff}}{U_\nu} = diag({m_{\nu _1}^2},{m_{\nu _2}^2},{m_{\nu _3}^2})\:,
\label{neutrino-diagonalize}
\end{eqnarray}
where the neutrino mass $m_{\nu _i}$ connected with experimental measurements through
\begin{eqnarray}
 {m_{\nu_2}^2 - m_{\nu_1}^2}  =  {\Delta m_{21}^2} , \qquad
 {m_{\nu_3}^2 - m_{\nu_2}^2}  =  {\Delta m_{32}^2} .
\label{mass-squared-neutrino}
\end{eqnarray}
The combination of Eq.(\ref{PMNS-matrix}), Eq.(\ref{neutrino-diagonalize}), Eq.(\ref{mass-squared-neutrino}) with neutrino oscillation experimental data gives some strong constraints on relevant parameter space of the $\mu\nu$SSM.

At the EW scale, the soft masses $m_{\tilde H_d}^2$, $m_{\tilde H_u}^2$, $m_{\tilde L_i}^2$ and $m_{\tilde \nu_i^c}^2$ are derived from the minimization conditions of the tree-level neutral scalar potential, which are given in \ref{appendix-mini}. Implying the approximate GUT relation $M_1=\frac{\alpha_1^2}{\alpha_2^2}M_2\approx 0.5 M_2$, the free parameters affect our analysis are
\begin{eqnarray}
\lambda , \: \kappa ,\: \tan \beta , \: {A_{\lambda,\kappa,\nu,e}},\:m_{{{\tilde e}^c}}, \: {\upsilon_{\nu^c}}, \:M_2\;.
\end{eqnarray}

To obtain the Yukawa couplings $Y_{\nu_i}$ and $\upsilon_{\nu_i}$ from Eq.(\ref{neutrino-diagonalize}),
we assume the neutrinos masses satisfying ${m_{\nu_1}}{\rm{ < }}{m_{\nu_2}}{\rm{ < }}{m_{\nu_3}}$,
and choose $m_{\nu_2}=10^{-2}\:{\rm{eV}}$ as input in our numerical analysis. Then we can get $m_{\nu_{1,3}}$ from the experimental data on the differences of neutrino mass squared. For $U_\nu$, the values of $\theta_{ij}$ are obtained from the experimental data in Eq.(\ref{neutrino-oscillations1}). And the effective light neutrino mass matrix $m_{eff}$ can approximate as \cite{Roy}
\begin{eqnarray}
{m_{ef{f_{ij}}}} \approx \frac{{2A{\upsilon_{\nu^c}}}}{{3\Delta }}{b_i}{b_j} + \frac{{1 - 3{\delta _{ij}}}}{{6\kappa {\upsilon_{\nu^c}}}}{a_i}{a_j}\:,
\end{eqnarray}
where
\begin{eqnarray}
&&\Delta  = {\lambda ^2}{(\upsilon_d^2 + \upsilon_u^2)}^2 + 4\lambda \kappa {\upsilon_{\nu^c}^2}{\upsilon_d}{\upsilon_u} - 12{\lambda ^2}{\upsilon_{\nu^c}}AB\:,\nonumber\\
&& A = \kappa {\upsilon_{\nu^c}^2} + \lambda {\upsilon_d}{\upsilon_u}\:,\nonumber\\
&& \frac{1}{B}= \frac{e^2}{c_{_W}^2{M_1}} + \frac{e^2}{s_{_W}^2{M_2}} \:, \nonumber\\
&&{a_i} = {Y_{{\nu _i}}}{\upsilon_u}\:, \quad {b_i} = {Y_{{\nu _i}}}{\upsilon_d} + 3\lambda \upsilon_{\nu_i}\:.
\end{eqnarray}
Then, we can numerically derive
$Y_{\nu_i} \sim \mathcal{O}(10^{-7})$ and $\upsilon_{\nu_i} \sim \mathcal{O}(10^{-4}{\rm{GeV}})$ from Eq.(\ref{neutrino-diagonalize}).

\subsection{Branching ratio of LFV processes}
\begin{figure}[htbp]
\setlength{\unitlength}{1mm}
\centering
\begin{minipage}[c]{0.7\textwidth}
\includegraphics[width=3.3in]{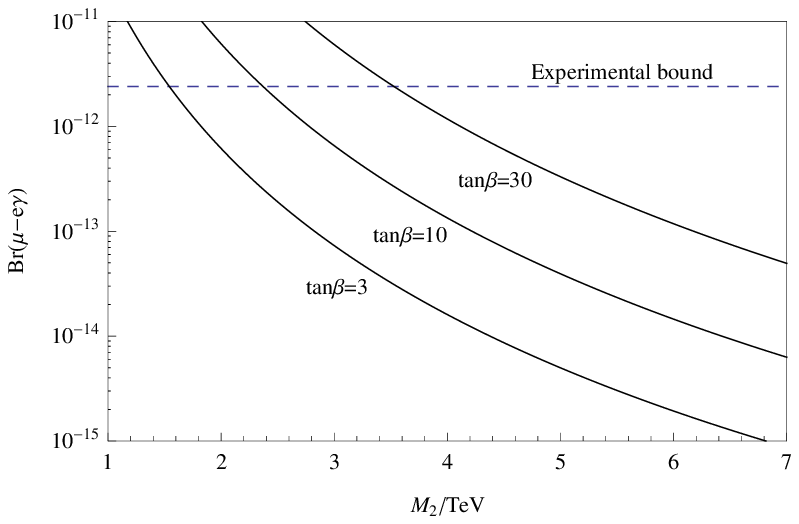}
\end{minipage}
\caption[]{Branching ratio for the process $\mu\rightarrow e\gamma$ varies with $M_2$
for $\tan\beta=3,\;10,\;30$, respectively. }
\label{fig4}
\end{figure}

\begin{figure}[htbp]
\setlength{\unitlength}{1mm}
\centering
\begin{minipage}[c]{0.7\textwidth}
\includegraphics[width=3.3in]{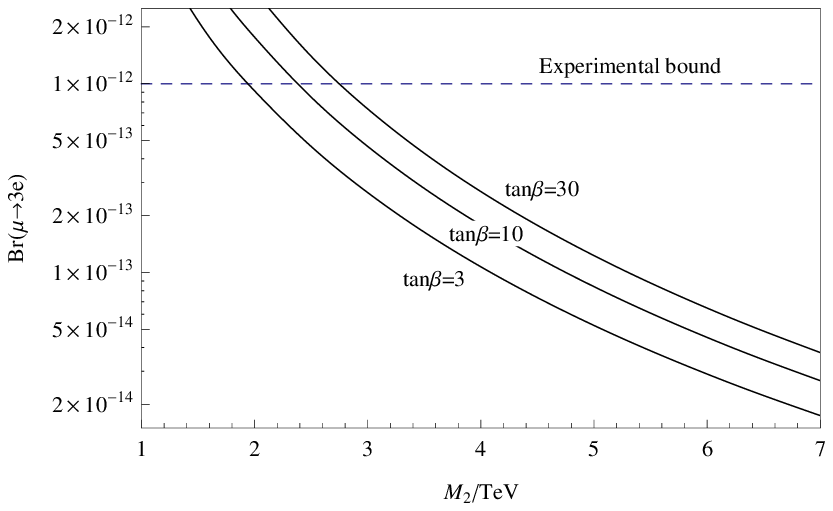}
\end{minipage}
\caption[]{Branching ratio for the process $\mu\rightarrow3e$  varies with $M_2$ for $\tan\beta=3,\;10,\;30$, respectively.}
\label{fig5}
\end{figure}

Considering the research of the $\mu\nu{\rm SSM}$ \cite{mnSSM}, we choose the relevant parameters as $\lambda=0.1$, $\kappa=0.01$, $m_{{{\tilde e}^c}}=A_e=A_\lambda=1\:{\rm{TeV}}$, $A_\nu=A_\kappa=-1\:{\rm{TeV}}$ and $\upsilon_{\nu^c}=800\:{\rm{GeV}}$ in next numerical analysis for convenience. With those
assumptions on parameter space, we present the branching ratio of $\mu\rightarrow e\gamma$ versus $M_2$ in Fig.\ref{fig4}.
As $M_2 \le 2\:{\rm{TeV}}$, the theoretical evaluations exceed the upper experimental bound easily. The fact implies that experimental data do not favor small $M_2$. Along with increasing of $M_2$, theoretical evaluation on the branching ratio of $\mu\rightarrow e\gamma$ decreases steeply. As $M_2=3\:{\rm{TeV}}$
and $\tan\beta=10$, theoretical evaluation on the branching ratio of $\mu\rightarrow e\gamma$ is about $5\times10^{-13}$ which can be detected in near future. In the future, the expected sensitivity for $ {\rm{Br}}(\mu\rightarrow e\gamma)$ would be of order $10^{-13}$ \cite{MEG}. Differing from LFV processes which are researched in the BRpV model \cite{BRpV-LFV}, the large VEVs of right-handed sneutrinos in the $\mu\nu$SSM induce new sources for lepton-flavor violation. So, here the branching ratio of $\mu\rightarrow e\gamma$ can easily reach the upper experimental bound $2.4\times10^{-12}$ \cite{PDG}.

We also investigate the $\mu\rightarrow3e$ processes in detail. And the branching ratio of $\mu\rightarrow3e$ is also decreases with increasing of $M_2$, and raises with increasing of $\tan\beta$, which is presented in the Fig.\ref{fig5}. By Introducing the right-handed sneutrinos which the VEVs are nonzero to the $\mu\nu$SSM, the branching ratio of $\mu\rightarrow3e$ can also easily reach the upper experimental bound $10^{-12}$ \cite{PDG}. We can see that the experimental bounds of the branching ratio of $\mu\rightarrow3e$ and  $\mu\rightarrow e\gamma$ give very strong constraints on the $\mu\nu{\rm SSM}$.

\begin{figure}[htbp]
\setlength{\unitlength}{1mm}
\centering
\begin{minipage}[c]{0.7\textwidth}
\includegraphics[width=3.3in]{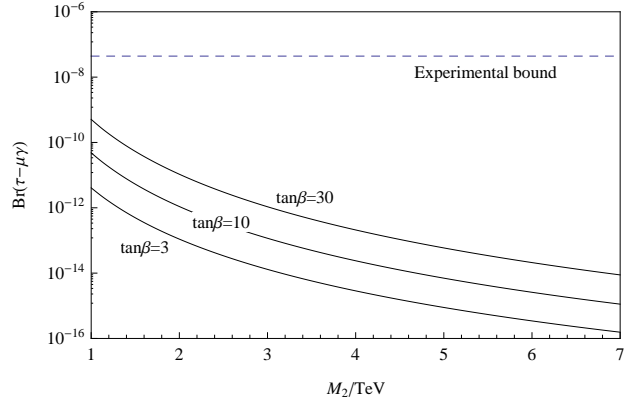}
\end{minipage}
\caption[]{Branching ratio for the process $\tau\rightarrow\mu\gamma$ varies with $M_2$
for $\tan\beta=3,\;10,\;30$, respectively.}
\label{fig6}
\end{figure}

In Fig.\ref{fig6}, we show the branching ratio for $\tau\rightarrow\mu\gamma$ versus $M_2$ as $\tan\beta=3,\;10,\;30$. Similar to the case of $\mu\rightarrow e\gamma$, the evaluation on the branching ratio for $\tau\rightarrow\mu\gamma$ decreases with increasing of $M_2$, and is enhanced by large $\tan\beta$. As $M_2=3\:{\rm TeV}$ and $\tan\beta=10$, ${\rm{Br}}(\tau\rightarrow\mu\gamma)\approx 10^{-13}$ is four orders below the expected sensitivity $10^{-9}$ \cite{Bona}.

\subsection{Muon anomalous magnetic dipole moment}
\begin{figure}[htbp]
\setlength{\unitlength}{1mm}
\centering
\begin{minipage}[c]{0.7\textwidth}
\includegraphics[width=3.5in]{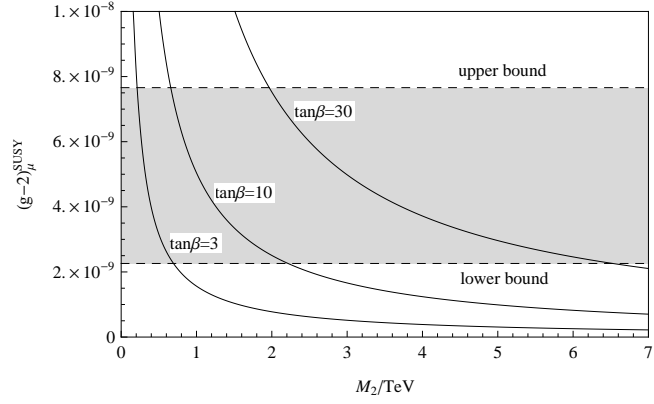}
\end{minipage}
\caption[]{The SUSY contribution to the anomalous magnetic dipole moment of the muon
varies with $M_2$ for $\tan\beta=3,\;10,\;30$, respectively. The gray area
 denotes the $\Delta a_\mu$ at $1.8$ standard deviation.}
\label{fig7}
\end{figure}

Finally, we analyze the anomalous magnetic dipole moment of the muon in the $\mu\nu$SSM. Rescaled the final result of the E821 Collaboration at BNL \cite{E821} using $\mu/p$ magnetic moment ratio of $3.183345137(85)$ from ref.\cite{Mohr}, the PDG Collaboration \cite{PDG} gives the world average of muon anomalous magnetic dipole moment
\begin{eqnarray}
a_\mu^{{\rm{exp}}} =\frac{1}{2}(g_\mu -2) = 11659208.9(5.4)(3.3)\times 10^{-10}\:,
\end{eqnarray}
where the statistical and systematic uncertainties are given, respectively. And the Standard Model (SM) prediction \cite{PDG} is
\begin{eqnarray}
a_\mu^{{\rm{SM}}} = 11659184.1(4.8)\times 10^{-10}\:.
\end{eqnarray}
So, the difference between experiment and the SM prediction
\begin{eqnarray}
\Delta a_\mu =a_\mu^{{\rm{exp}}} -a_\mu^{{\rm{SM}}} = 24.8(8.7)(4.8)\times 10^{-10}\:,
\end{eqnarray}
represents an interesting but not yet conclusive discrepancy of $1.8$ standard deviation.
An alternate interpretation is that $\Delta a_\mu$ may be a new physics signal with supersymmetric
particle loops as the leading candidate explanation. If treated the supersymmetry as the leading explanation, parameter space of the $\mu\nu$SSM should be constrained by the experimental data on $\Delta a_\mu$.

The SUSY contribution to the muon anomalous magnetic dipole moment
in the $\mu\nu$SSM is shown in Fig.\ref{fig7}. The result shows that when $\tan\beta=3$, $\Delta a_\mu$ constrains $M_2<1\:{\rm{TeV}}$, which is opposite to what the upper experimental bound of $\rm{Br}(\mu\rightarrow e\gamma)$ constrains.  The fact implies that experimental
data do not favor small $\tan\beta$ in the $\mu\nu$SSM with the MFV assumptions (\ref{MFV}). When $\tan\beta=30$, $\Delta a_\mu$ constrains $2\:{\rm{TeV}}\leq M_2<7\:{\rm{TeV}}$, compared with that the upper experimental bound of $\rm{Br}(\mu\rightarrow e\gamma)$ constrains $ M_2\geq 3.5\:{\rm{TeV}}$, the $M_2$ has more consistent interval. So, under the MFV assumptions,
the $\mu\nu$SSM favors large $\tan\beta$ and $M_2$ for consistent with
experimental data.

\section{Conclusions\label{sec:6}}
Besides the superfields of the MSSM, the $\mu\nu$SSM introduces three exotic right-handed sneutrinos $\hat{\nu}_i^c$ to solve the $\mu$ problem of the MSSM. And exotic right-handed sneutrinos which the vacuum expectation values are nonzero induce new sources for lepton-flavor violation. In addition, from the scalars for the $\mu\nu$SSM we strictly separate the Goldstone bosons, which disappear in the physical gauge.

Considering the updated experimental data on neutrino oscillations, we analyze various LFV processes and $(g-2)_\mu$ in the $\mu\nu$SSM. Numerical results indicate that the new physics corrections dominate the evaluation on the branching ratios of LFV processes in some parameter space of the $\mu\nu$SSM. And the theoretical predictions on the branching ratios of LFV processes $\mu\rightarrow e\gamma$ and $\mu\rightarrow3e$ for large $\tan\beta$ can easily reach the present experimental upper bounds and be detected in near future. Additionally, the present experimental observations on $(g-2)_\mu$ also give very strong constraint on the model. Under the MFV assumptions (\ref{MFV}), the $\mu\nu$SSM favors large $\tan\beta$ and $M_2$ for consistent with experimental data. Certainly, a neutral Higgs with mass $m_{h_0}\sim 124-126\;{\rm GeV}$ reported by ATLAS \cite{ATLAS} and CMS \cite{CMS} also contributes a strict constraint on relevant parameter space, we will discuss this problem elsewhere.

\section*{Acknowledgements}

The work has been supported by the National Natural Science Foundation of China (NNSFC)
with Grant No. 10975027, No. 11275036, No. 11047002 and Natural Science Fund of Hebei University
with Grant No. 2011JQ05, No. 2012-242.

\appendix

\section{Minimization of the potential\label{appendix-mini}}

First, the eight minimization conditions of the tree-level neutral scalar potential are given below:
\begin{eqnarray}
&&m_{{H_d}}^2 \upsilon_d + \frac{G^2}{4}( \upsilon_d^2 - \upsilon_u^2 +\upsilon_{\nu_i}\upsilon_{\nu_i}) \upsilon_d  - (A_\lambda \lambda)_i {\upsilon_u} \upsilon_{\nu_i^c} - {\lambda _j}{\kappa _{ijk}}{\upsilon_u}\upsilon_{\nu_i^c} \upsilon_{\nu_k^c}  \nonumber\\
&&\qquad  + \:({\lambda _i}{\lambda _j}\upsilon_{\nu_i^c}\upsilon_{\nu_j^c}  + {\lambda _i}{\lambda _i}\upsilon_u^2){\upsilon_d}  - {Y_{{\nu_{ij}}}}\upsilon_{\nu_i}({\lambda _k}\upsilon_{\nu_k^c}\upsilon_{\nu_j^c} + {\lambda _j}\upsilon_u^2) = 0 \:,\\
&&m_{{H_u}}^2{\upsilon_u} - \frac{{G^2}}{4}(\upsilon_d^2 - \upsilon_u^2 +\upsilon_{\nu_i}\upsilon_{\nu_i})\upsilon_u  + {(A_\nu Y_\nu)}_{ij}\upsilon_{\nu_i}\upsilon_{\nu_j^c} - (A_\lambda \lambda)_i {\upsilon_d} \upsilon_{\nu_i^c}  \nonumber\\
&&\qquad  + \:({\lambda _i}{\lambda _j}\upsilon_{\nu_i^c}\upsilon_{\nu_j^c}  + {\lambda _i}{\lambda _i}\upsilon_u^2){\upsilon_u} + {Y_{{\nu_{ij}}}}\upsilon_{\nu_i}({\kappa _{ljk}}\upsilon_{\nu_l^c} \upsilon_{\nu_k^c} - 2 {\lambda _j}\upsilon_d \upsilon_u )  \nonumber\\
&&\qquad - \: {\lambda _j}{\kappa _{ijk}}{\upsilon_d}\upsilon_{\nu_i^c} \upsilon_{\nu_k^c} +  ( Y_{\nu_{ki}}Y_{\nu_{kj}}\upsilon_{\nu_i^c}\upsilon_{\nu_j^c}  + Y_{\nu_{ik}}Y_{\nu_{jk}}\upsilon_{\nu_i}\upsilon_{\nu_j} )  \upsilon_u  = 0 \:,\\
&&m_{{{\tilde L}_{ij}}}^2\upsilon_{\nu_j} + \frac{{G^2}}{4}(\upsilon_d^2 - \upsilon_u^2 +\upsilon_{\nu_j}\upsilon_{\nu_j})\upsilon_{\nu_i}   + {(A_\nu Y_\nu)}_{ij}\upsilon_u\upsilon_{\nu_j^c} + {Y_{{\nu_{il}}}}{\kappa _{ljk}}\upsilon_u \upsilon_{\nu_j^c} \upsilon_{\nu_k^c}   \quad \nonumber\\
&&\qquad - \:  {Y_{{\nu_{ij}}}}{\lambda _k}\upsilon_{\nu_j^c} \upsilon_{\nu_k^c}  \upsilon_d -{Y_{{\nu_{ij}}}}{\lambda _j}\upsilon_u^2  \upsilon_d + {Y_{{\nu_{ij}}}}{Y_{{\nu_{lk}}}}\upsilon_{\nu_l}\upsilon_{\nu_j^c} \upsilon_{\nu_k^c}\nonumber\\
&&\qquad + \: {Y_{{\nu_{ik}}}}{Y_{{\nu_{jk}}}}\upsilon_u^2\upsilon_{\nu_j} = 0 \:,\\
&&m_{\tilde \nu_{ij}^c}^2 \upsilon_{\nu_j^c} + {(A_\nu Y_\nu)}_{ji}\upsilon_{\nu_j}\upsilon_u - (A_\lambda \lambda)_i{\upsilon_d}{\upsilon_u} +{( A_\kappa \kappa)}_{ijk} \upsilon_{\nu_j^c} \upsilon_{\nu_k^c}  -  2{\lambda _j}{\kappa _{ijk}}{\upsilon_d}{\upsilon_u}\upsilon_{\nu_k^c} \nonumber\\
&&\qquad + \: {\lambda _i}{\lambda _j}\upsilon_{\nu_j^c}(\upsilon_d^2  + \upsilon_u^2) + 2{\kappa _{lim}}{\kappa _{ljk}} \upsilon_{\nu_m^c} \upsilon_{\nu_j^c} \upsilon_{\nu_k^c} +  2{Y_{{\nu_{jk}}}}{\kappa _{ikl}}{\upsilon_u}\upsilon_{\nu_j} \upsilon_{\nu_l^c} \nonumber\\
&&\qquad -\: {Y_{{\nu_{ji}}}}{\lambda _k}\upsilon_{\nu_j} \upsilon_{\nu_k^c}{\upsilon_d} - {Y_{{\nu_{kj}}}}{\lambda _i}\upsilon_{\nu_k} \upsilon_{\nu_j^c} {\upsilon_d} + {Y_{{\nu_{ji}}}}{Y_{{\nu_{lk}}}}\upsilon_{\nu_j}\upsilon_{\nu_l}\upsilon_{\nu_k^c} \nonumber\\
&&\qquad +\: {Y_{{\nu_{ki}}}}{Y_{{\nu_{kj}}}}\upsilon_u^2\upsilon_{\nu_j^c} = 0 \:,
\end{eqnarray}
where $G^2=g_1^2+g_2^2$ and $g_1 c_{_W} =g_2 s_{_W}=e$.

\section{Mass Matrices\label{appendix-mass}}
In this appendix, we give the mass matrices in the $\mu\nu$SSM.

\subsection{Scalar mass matrices}
For this subsection, we use the indices $i,j,k,l,m=1,2,3$ and $\alpha=1,\ldots,8$.

\subsubsection{CP-even neutral scalars}
In the unrotated basis ${S'^T} = ({h_d},{h_u},{(\tilde \nu_i)^R},{({\tilde \nu_i^c})^R})$, one can obtain the quadratic potential
\begin{eqnarray}
{V_{quadratic}} = \frac{1}{2} {S'^T}M_S^2S' \: .
\end{eqnarray}
And the expression for the independent coefficients of $M_S^2$ are given in detail below:
\begin{eqnarray}
&&M_{h_d h_d}^2 = m_{H_d}^2 + \frac{G^2}{4}(3\upsilon_d^2-\upsilon_u^2+\upsilon_{\nu_i}\upsilon_{\nu_i})+\lambda_i \lambda_j \upsilon_{\nu_i^c}\upsilon_{\nu_j^c}+\lambda_i \lambda_i \upsilon_u^2 \: , \\
&&M_{h_u h_u}^2 = m_{H_u}^2 - \frac{G^2}{4}(\upsilon_d^2 - 3\upsilon_u^2 + \upsilon_{\nu_i}\upsilon_{\nu_i})+\lambda_i \lambda_j \upsilon_{\nu_i^c}\upsilon_{\nu_j^c}+\lambda_i \lambda_i \upsilon_d^2 \nonumber\\
&&\qquad\qquad - \: 2Y_{\nu_{ij}}\lambda_j \upsilon_d \upsilon_{\nu_i} + Y_{\nu_{ki}}Y_{\nu_{kj}}\upsilon_{\nu_i^c}\upsilon_{\nu_j^c}  + Y_{\nu_{ik}}Y_{\nu_{jk}}\upsilon_{\nu_i}\upsilon_{\nu_j} \:, \\
&&M_{h_d h_u}^2 = -(A_\lambda \lambda)_i \upsilon_{\nu_i^c} - \frac{G^2}{2}\upsilon_d \upsilon_u +2 \lambda_i \lambda_i \upsilon_d \upsilon_u  -\lambda_k \kappa_{ijk}\upsilon_{\nu_i^c}\upsilon_{\nu_j^c} \nonumber\\
&&\qquad\qquad - \: 2Y_{\nu_{ij}} \lambda_j \upsilon_u \upsilon_{\nu_i} \:, \\
&&M_{h_d (\tilde \nu_i)^R}^2 = \frac{G^2}{2}\upsilon_d \upsilon_{\nu_i} - Y_{\nu_{ij}}(\lambda_j \upsilon_u^2 + \lambda_k \upsilon_{\nu_k^c} \upsilon_{\nu_j^c}) \: ,\\
&&M_{h_u (\tilde \nu_i)^R}^2 = - \frac{G^2}{2}\upsilon_u \upsilon_{\nu_i} + {(A_\nu Y_\nu)}_{ij} \upsilon_{\nu_j^c} - 2 Y_{\nu_{ij}}\lambda_j \upsilon_d \upsilon_u + Y_{\nu_{ik}} \kappa_{ljk} \upsilon_{\nu_l^c}\upsilon_{\nu_j^c} \nonumber\\
&&\qquad\qquad\quad +\: 2 Y_{\nu_{ij}} Y_{\nu_{kj}}  \upsilon_u \upsilon_{\nu_k} \:, \\
&&M_{h_d (\tilde \nu_i^c)^R}^2 = -(A_\lambda \lambda)_i \upsilon_u +2\lambda_i \lambda_j \upsilon_d \upsilon_{\nu_j^c} - 2 \lambda_k \kappa_{ijk}\upsilon_u \upsilon_{\nu_j^c} \nonumber\\
&&\qquad\qquad\quad - \:( Y_{\nu_{ji}} \lambda_k  + Y_{\nu_{jk}} \lambda_i ) \upsilon_{\nu_j} \upsilon_{\nu_k^c} \:, \\
&&M_{h_u (\tilde \nu_i^c)^R}^2 = -(A_\lambda \lambda)_i \upsilon_d + {(A_\nu Y_\nu)}_{ji} \upsilon_{\nu_j} +2\lambda_i \lambda_j \upsilon_u\upsilon_{\nu_j^c} - 2 \lambda_k \kappa_{ijk}\upsilon_d \upsilon_{\nu_j^c} \qquad\;\;\; \nonumber\\
&&\qquad\qquad\quad + \: 2Y_{\nu_{jk}} \kappa_{ilk} \upsilon_{\nu_j}\upsilon_{\nu_l^c} +2Y_{\nu_{jk}} Y_{\nu_{ji}}  \upsilon_u \upsilon_{\nu_k^c} \:, \\
&&M_{(\tilde \nu_i)^R (\tilde \nu_j)^R}^2 = m_{\tilde L_{ij}}^2 +  \frac{G^2}{2}\upsilon_{\nu_i} \upsilon_{\nu_j} + \frac{G^2}{4}(\upsilon_d^2 - \upsilon_u^2 + \upsilon_{\nu_k}\upsilon_{\nu_k})\delta_{ij} \nonumber\\
&&\qquad\qquad\quad\;\; +\: Y_{\nu_{ik}} Y_{\nu_{jk}}  \upsilon_u^2 + Y_{\nu_{ik}} Y_{\nu_{jl}}  \upsilon_{\nu_k^c}\upsilon_{\nu_l^c} \:, \\
&&M_{(\tilde \nu_i)^R (\tilde \nu_j^c)^R}^2 = {(A_\nu Y_\nu)}_{ij}\upsilon_u - (Y_{\nu_{ij}} \lambda_k + Y_{\nu_{ik}}\lambda_j ) \upsilon_d \upsilon_{\nu_k^c} + 2 Y_{\nu_{ik}} \kappa_{jlk} \upsilon_u \upsilon_{\nu_l^c}
\nonumber\\
&&\qquad\qquad\quad\;\; +\:(Y_{\nu_{ij}} Y_{\nu_{kl}} +Y_{\nu_{il}} Y_{\nu_{kj}} ) \upsilon_{\nu_k} \upsilon_{\nu_l^c}  \:, \\
&&M_{(\tilde \nu_i^c)^R (\tilde \nu_j^c)^R}^2 =  m_{\tilde \nu_{ij}^c}^2 + 2 {(A_\kappa \kappa)}_{ijk} \upsilon_{\nu_k^c} - 2\lambda_k \kappa_{ijk} \upsilon_d \upsilon_u + \lambda_i \lambda_j ( \upsilon_d^2 + \upsilon_u^2) \nonumber\\
&&\qquad\qquad\quad\;\; +\:(2\kappa_{ijk}\kappa_{lmk}+4\kappa_{ilk}\kappa_{jmk}) \upsilon_{\nu_l^c}\upsilon_{\nu_m^c} + 2Y_{\nu_{lk}} \kappa_{ijk} \upsilon_u \upsilon_{\nu_l} \nonumber\\
&&\qquad\qquad\quad\;\; -\: (Y_{\nu_{kj}}\lambda_i + Y_{\nu_{ki}}\lambda_j)\upsilon_d \upsilon_{\nu_k} +Y_{\nu_{ki}}(Y_{\nu_{kj}} \upsilon_u^2 + Y_{\nu_{lj}}\upsilon_{\nu_k}\upsilon_{\nu_l}) \:.
\end{eqnarray}

We can use an $8\times8$ unitary matrix $R_S$ to diagonalize the mass matrix $M_S^2$
\begin{eqnarray}
R_S^TM_S^2{R_S} = {(M_S^{diag})^2}\:.
\end{eqnarray}
By unitary matrix $R_S$, $S'_\alpha$ can be rotated to the mass eigenvectors $S_\alpha$:
\begin{eqnarray}
{h_d} = R_S^{1\alpha }{S_\alpha },\; {h_u} = R_S^{2\alpha }{S_\alpha },\; {(\tilde \nu_i)^R} = R_S^{(2 + i)\alpha }{S_\alpha },\; {({\tilde \nu_i^c})^R} = R_S^{(5 + i)\alpha }{S_\alpha }\:.
\end{eqnarray}

\subsubsection{CP-odd neutral scalars}
In the unrotated basis ${P'^T} = ({P_d},{P_u},{(\tilde \nu_i)^I},{({\tilde \nu_i^c})^I})$, one can also give the quadratic potential
\begin{eqnarray}
{V_{quadratic}} = \frac{1}{2}{P'^T}M_P^2P' \: ,
\end{eqnarray}
and the concrete expression for the independent coefficients of $M_P^2$
\begin{eqnarray}
&&M_{P_d P_d}^2 = m_{H_d}^2 + \frac{G^2}{4}(\upsilon_d^2-\upsilon_u^2+\upsilon_{\nu_i}\upsilon_{\nu_i})+\lambda_i \lambda_j \upsilon_{\nu_i^c}\upsilon_{\nu_j^c}+\lambda_i \lambda_i \upsilon_u^2 \: , \\
&&M_{P_u P_u}^2 = m_{H_u}^2 - \frac{G^2}{4}(\upsilon_d^2 - \upsilon_u^2 + \upsilon_{\nu_i}\upsilon_{\nu_i})+\lambda_i \lambda_j \upsilon_{\nu_i^c}\upsilon_{\nu_j^c}+\lambda_i \lambda_i \upsilon_d^2 \qquad\qquad\quad\; \nonumber\\
&&\qquad\qquad - \: 2Y_{\nu_{ij}}\lambda_j \upsilon_d \upsilon_{\nu_i} + Y_{\nu_{ki}}Y_{\nu_{kj}}\upsilon_{\nu_i^c}\upsilon_{\nu_j^c}  + Y_{\nu_{ik}}Y_{\nu_{jk}}\upsilon_{\nu_i}\upsilon_{\nu_j} \:, \\
&&M_{P_d P_u}^2 = (A_\lambda \lambda)_i \upsilon_{\nu_i^c} + \lambda_k \kappa_{ijk}\upsilon_{\nu_i^c}\upsilon_{\nu_j^c}  \:, \\
&&M_{P_d (\tilde \nu_i)^I}^2 =  - Y_{\nu_{ij}}(\lambda_j \upsilon_u^2 + \lambda_k \upsilon_{\nu_k^c} \upsilon_{\nu_j^c}) \: ,\\
&&M_{P_u (\tilde \nu_i)^I}^2 = - {(A_\nu Y_\nu)}_{ij} \upsilon_{\nu_j^c} -  Y_{\nu_{ik}} \kappa_{ljk} \upsilon_{\nu_l^c}\upsilon_{\nu_j^c}  \:, \\
&&M_{P_d (\tilde \nu_i^c)^I}^2 = (A_\lambda \lambda)_i \upsilon_u  - 2 \lambda_k \kappa_{ijk}\upsilon_u \upsilon_{\nu_j^c}  - ( Y_{\nu_{ji}} \lambda_k  - Y_{\nu_{jk}} \lambda_i ) \upsilon_{\nu_j} \upsilon_{\nu_k^c} \:, \\
&&M_{P_u (\tilde \nu_i^c)^I}^2 = (A_\lambda \lambda)_i \upsilon_d - {(A_\nu Y_\nu)}_{ji} \upsilon_{\nu_j} - 2( \lambda_k \kappa_{ilk}\upsilon_d   -  Y_{\nu_{jk}} \kappa_{ilk} \upsilon_{\nu_j})\upsilon_{\nu_l^c} \:, \\
&&M_{(\tilde \nu_i)^I (\tilde \nu_j)^I}^2 = m_{\tilde L_{ij}}^2 + \frac{G^2}{4}(\upsilon_d^2 - \upsilon_u^2 + \upsilon_{\nu_k}\upsilon_{\nu_k})\delta_{ij}  + Y_{\nu_{ik}} Y_{\nu_{jk}}  \upsilon_u^2 \nonumber\\
&&\qquad\qquad\quad\;\; + \: Y_{\nu_{ik}} Y_{\nu_{jl}}  \upsilon_{\nu_k^c}\upsilon_{\nu_l^c} \:, \\
&&M_{(\tilde \nu_i)^I (\tilde \nu_j^c)^I}^2 = -{(A_\nu Y_\nu)}_{ij}\upsilon_u + (Y_{\nu_{ij}} \lambda_k - Y_{\nu_{ik}}\lambda_j ) \upsilon_d \upsilon_{\nu_k^c} + 2 Y_{\nu_{il}} \kappa_{jlk} \upsilon_u \upsilon_{\nu_k^c}
\nonumber\\
&&\qquad\qquad\quad\;\; -\:(Y_{\nu_{ij}} Y_{\nu_{kl}} - Y_{\nu_{il}} Y_{\nu_{kj}} ) \upsilon_{\nu_k} \upsilon_{\nu_l^c}  \:, \\
&&M_{(\tilde \nu_i^c)^I (\tilde \nu_j^c)^I}^2 =  m_{\tilde \nu_{ij}^c}^2 - 2 {(A_\kappa \kappa)}_{ijk} \upsilon_{\nu_k^c} + 2\lambda_k \kappa_{ijk} \upsilon_d \upsilon_u + \lambda_i \lambda_j ( \upsilon_d^2 + \upsilon_u^2) \nonumber\\
&&\qquad\qquad\quad\;\; -\:(2\kappa_{ijk}\kappa_{lmk}-4\kappa_{imk}\kappa_{ljk}) \upsilon_{\nu_l^c}\upsilon_{\nu_m^c} - 2Y_{\nu_{lk}} \kappa_{ijk} \upsilon_u \upsilon_{\nu_l} \nonumber\\
&&\qquad\qquad\quad\;\; -\: (Y_{\nu_{kj}}\lambda_i + Y_{\nu_{ki}}\lambda_j)\upsilon_d \upsilon_{\nu_k} +Y_{\nu_{ki}}(Y_{\nu_{kj}} \upsilon_u^2 + Y_{\nu_{lj}}\upsilon_{\nu_k}\upsilon_{\nu_l}) \:.
\end{eqnarray}

Using an $8\times8$ unitary matrix $R_P$ to diagonalize the mass matrix $M_P^2$
\begin{eqnarray}
R_P^TM_P^2{R_P} = {(M_P^{diag})^2}\:,
\end{eqnarray}
we can obtain the mass eigenvectors $P_\alpha$:
\begin{eqnarray}
{P_d} = R_P^{1\alpha }{P_\alpha },\; {P_u} = R_P^{2\alpha }{P_\alpha },\; {(\tilde \nu_i)^I} = R_P^{(2 + i)\alpha }{P_\alpha },\; {({\tilde \nu_i^c})^I} = R_P^{(5 + i)\alpha }{P_\alpha }\:.
\end{eqnarray}

\subsubsection{Charged scalars}
The quadratic potential includes
\begin{eqnarray}
{V_{quadratic}} = {S'^{-T}}M_{S^\pm}^2S'^+ \: ,
\end{eqnarray}
where ${S'^{ \pm T}} = (H_d^ \pm ,H_u^ \pm ,\tilde e_{L_i}^ \pm ,\tilde e_{R_i}^ \pm )$ is in the unrotated basis, $\tilde e_{L_i}^- \equiv \tilde e_i$ and $\tilde e_{R_i}^+ \equiv \tilde e_i^c$. The concrete expression for the independent coefficients of $M_{S^\pm}^2$ are given below:
\begin{eqnarray}
&&M_{H_d^\pm H_d^\pm}^2 = m_{H_d}^2 + \frac{g_2^2}{2}(\upsilon_u^2-\upsilon_{\nu_i}\upsilon_{\nu_i}) + \frac{G^2}{4}(\upsilon_d^2-\upsilon_u^2+\upsilon_{\nu_i}\upsilon_{\nu_i})+\lambda_i \lambda_j \upsilon_{\nu_i^c}\upsilon_{\nu_j^c} \quad \nonumber\\
&&\qquad\qquad\;\; + \: Y_{e_{ik}}Y_{e_{jk}}\upsilon_{\nu_i}\upsilon_{\nu_j} \: , \\
&&M_{H_u^\pm H_u^\pm}^2 = m_{H_u}^2 + \frac{g_2^2}{2}(\upsilon_d^2+\upsilon_{\nu_i}\upsilon_{\nu_i}) - \frac{G^2}{4}(\upsilon_d^2 - \upsilon_u^2 + \upsilon_{\nu_i}\upsilon_{\nu_i})+\lambda_i \lambda_j \upsilon_{\nu_i^c}\upsilon_{\nu_j^c} \nonumber\\
&&\qquad\qquad\;\; + \: Y_{\nu_{ik}}Y_{\nu_{ij}}\upsilon_{\nu_j^c}\upsilon_{\nu_k^c}  \:, \\
&&M_{H_d^\pm H_u^\pm}^2 = (A_\lambda \lambda)_i \upsilon_{\nu_i^c} + \frac{g_2^2}{2}\upsilon_d \upsilon_u - \lambda_i \lambda_i \upsilon_d \upsilon_u  +\lambda_k \kappa_{ijk}\upsilon_{\nu_i^c}\upsilon_{\nu_j^c} \nonumber\\
&&\qquad\qquad\;\; + \: Y_{\nu_{ij}} \lambda_j \upsilon_u \upsilon_{\nu_i} \:, \\
&&M_{H_d^\pm \tilde e_{L_i}^\pm}^2 = \frac{g_2^2}{2}\upsilon_d \upsilon_{\nu_i} - Y_{\nu_{ij}}\lambda_k \upsilon_{\nu_k^c} \upsilon_{\nu_j^c} -  Y_{e_{ij}}Y_{e_{kj}}\upsilon_d \upsilon_{\nu_k} \: ,\\
&&M_{H_u^\pm \tilde e_{L_i}^\pm}^2 = \frac{g_2^2}{2}\upsilon_u \upsilon_{\nu_i} -  {(A_\nu Y_\nu)}_{ij} \upsilon_{\nu_j^c} + Y_{\nu_{ij}}\lambda_j \upsilon_d \upsilon_u - Y_{\nu_{ij}} \kappa_{ljk} \upsilon_{\nu_l^c}\upsilon_{\nu_k^c} \nonumber\\
&&\qquad\qquad\;\; -\:  Y_{\nu_{ik}} Y_{\nu_{kj}}  \upsilon_u \upsilon_{\nu_j} \:, \\
&&M_{H_d^\pm \tilde e_{R_i}^\pm}^2 = -(A_e Y_e)_{ji} \upsilon_{\nu_j}  -  Y_{e_{ki}}Y_{\nu_{kj}}\upsilon_u \upsilon_{\nu_j^c} \:, \\
&&M_{H_u^\pm \tilde e_{R_i}^\pm}^2 = - Y_{e_{ki}} (\lambda_j \upsilon_{\nu_j^c}\upsilon_{\nu_k} +Y_{\nu_{kj}}  \upsilon_d \upsilon_{\nu_j^c}) \:, \\
&&M_{\tilde e_{L_i}^\pm \tilde e_{L_j}^\pm}^2 = m_{\tilde L_{ij}}^2 + \frac{1}{4}(g_1^2-g_2^2)(\upsilon_d^2 - \upsilon_u^2 + \upsilon_{\nu_k}\upsilon_{\nu_k})\delta_{ij} + \frac{g_2^2}{2}\upsilon_{\nu_i} \upsilon_{\nu_j}  \nonumber\\
&&\qquad\qquad\;\; +\: Y_{\nu_{il}} Y_{\nu_{jk}}  \upsilon_{\nu_l^c}\upsilon_{\nu_k^c} +  Y_{e_{ik}} Y_{e_{jk}}  \upsilon_d^2  \:, \\
&&M_{\tilde e_{L_i}^\pm \tilde e_{R_j}^\pm}^2 = {(A_e Y_e)}_{ij}\upsilon_d - Y_{e_{ij}} \lambda_k  \upsilon_u \upsilon_{\nu_k^c}   \:, \\
&&M_{\tilde e_{R_i}^\pm \tilde e_{R_j}^\pm}^2 =  m_{\tilde e_{ij}^c}^2 - \frac{1}{2}g_1^2(\upsilon_d^2 - \upsilon_u^2 + \upsilon_{\nu_k}\upsilon_{\nu_k})\delta_{ij}  +  Y_{e_{ki}}Y_{e_{kj}} \upsilon_d^2 \nonumber\\
&&\qquad\qquad\;\; +\: Y_{e_{li}}Y_{e_{kj}}\upsilon_{\nu_k}\upsilon_{\nu_l} \:.
\end{eqnarray}

Through an $8\times8$ unitary matrix $R_{S^\pm}$ to diagonalize the mass matrix $M_{S^\pm}^2$
\begin{eqnarray}
R_{S^\pm}^TM_{S^\pm}^2{R_{S^\pm}} = {(M_{S^\pm}^{diag})^2}\:,
\end{eqnarray}
$S'^ \pm _\alpha$ can be rotated to the mass eigenvectors $S^ \pm _\alpha $:
\begin{eqnarray}
H_d^ \pm  = R_{{S^ \pm }}^{1\alpha }S_\alpha ^ \pm ,\; H_u^ \pm = R_{{S^ \pm }}^{2\alpha }S_\alpha ^ \pm ,\; \tilde e_{L_i}^ \pm =R_{{S^ \pm }}^{(2 + i)\alpha }S_\alpha ^ \pm ,\; \tilde e_{R_i}^ \pm = R_{{S^ \pm }}^{(5 + i)\alpha }S_\alpha ^ \pm \:.
\end{eqnarray}

\subsection{Neutral fermion mass matrix}
Neutrinos mix with the neutralinos and therefore in the unrotated basis ${\chi '^{ \circ T}} = \left( {{{\tilde B}^ \circ },{{\tilde W}^ \circ },{{\tilde H}_d}{\rm{,}}{{\tilde H}_u},{\nu_{R_i}},{\nu_{L{_i}}}} \right)$, one can have the neutral fermion mass terms in the Lagrangian:
\begin{eqnarray}
 - \frac{1}{2}{\chi '^{ \circ T}}{M_n}{\chi '^ \circ } + {\rm{H.c.}}\:  ,
\end{eqnarray}
where
\begin{eqnarray}
{M_n} = \left( {\begin{array}{*{20}{c}}
   M & {{m^T}}  \\
   m & {{0_{3 \times 3}}}  \\
\end{array}} \right),
\end{eqnarray}
with
\begin{eqnarray}
m = \left( {\begin{array}{*{20}{c}}
   {  -\frac{g_1}{\sqrt 2 }\upsilon_{{\nu _1}}} & { \frac{g_2}{\sqrt 2 }\upsilon_{{\nu _1}}} & 0 & {{Y_{{\nu _{1i}}}}{\upsilon_{\nu _i^c}}} & {{Y_{{\nu _{11}}}}{\upsilon_u}} & {{Y_{{\nu _{12}}}}{\upsilon_u}} & {{Y_{{\nu _{13}}}}{\upsilon_u}}  \\
   {  -\frac{g_1}{\sqrt 2 }\upsilon_{{\nu _2}}} & { \frac{g_2}{\sqrt 2 }\upsilon_{{\nu _2}}} & 0 & {{Y_{{\nu _{2i}}}}{\upsilon_{\nu _i^c}}} & {{Y_{{\nu _{21}}}}{\upsilon_u}} & {{Y_{{\nu _{22}}}}{\upsilon_u}} & {{Y_{{\nu _{23}}}}{\upsilon_u}}  \\
   {  -\frac{g_1}{\sqrt 2 }\upsilon_{{\nu _3}}} & { \frac{g_2}{\sqrt 2 }\upsilon_{{\nu _3}}} & 0 & {{Y_{{\nu _{3i}}}}{\upsilon_{\nu _i^c}}} & {{Y_{{\nu _{31}}}}{\upsilon_u}} & {{Y_{{\nu _{32}}}}{\upsilon_u}} & {{Y_{{\nu _{33}}}}{\upsilon_u}}  \\
\end{array}} \right)
\end{eqnarray}
and
\begin{eqnarray}
&&M = \left( {\begin{array}{*{20}{c}}
   {{M_1}} & 0 & {\frac{-g_1}{{\sqrt 2 }}{\upsilon _d}} & {\frac{g_1}{{\sqrt 2 }}{\upsilon _u}} & 0 & 0 & 0  \\
   0 & {{M_2}} & {\frac{g_2}{{\sqrt 2 }}{\upsilon _d}} & {\frac{-g_2}{{\sqrt 2 }}{\upsilon _u}} & 0 & 0 & 0  \\
   {\frac{-g_1}{{\sqrt 2 }}{\upsilon _d}} & {\frac{g_2}{{\sqrt 2 }}{\upsilon _d}} & 0 & {-{\lambda _i}{\upsilon _{\nu _i^c}}} & { - {\lambda _1}{\upsilon _u}} & { - {\lambda _2}{\upsilon _u}} & { - {\lambda _3}{\upsilon _u}}  \\
   {\frac{g_1}{{\sqrt 2 }}{\upsilon _u}} & {\frac{-g_2}{{\sqrt 2 }}{\upsilon _u}} & {-{\lambda _i}{\upsilon _{\nu _i^c}}} & 0 & {y_1} & {y_2} & { y_3}  \\
   0 & 0 & { - {\lambda _1}{\upsilon _u}} & { y_1} & {2{\kappa _{11j}}{\upsilon _{\nu _j^c}}} & {2{\kappa _{12j}}{\upsilon _{\nu _j^c}}} & {2{\kappa _{13j}}{\upsilon _{\nu _j^c}}}  \\
   0 & 0 & { - {\lambda _2}{\upsilon _u}} & { y_2} & {2{\kappa _{21j}}{\upsilon _{\nu _j^c}}} & {2{\kappa _{22j}}{\upsilon _{\nu _j^c}}} & {2{\kappa _{23j}}{\upsilon _{\nu _j^c}}}  \\
   0 & 0 & { - {\lambda _3}{\upsilon _u}} & { y_3} & {2{\kappa _{31j}}{\upsilon _{\nu _j^c}}} & {2{\kappa _{32j}}{\upsilon _{\nu _j^c}}} & {2{\kappa _{33j}}{\upsilon _{\nu _j^c}}}  \\
\end{array}} \right) \nonumber\\
&&
\end{eqnarray}
where $y_i=- {\lambda _i}{\upsilon _d}+ {{Y_{{\nu _{ji}}}}{\upsilon _{{\nu _j}}} }$. Here, the submatrix $m$ is neutralino-neutrino mixing, and the submatrix $M$ is neutralino mass matrix. This $10\times10$ symmetric matrix $M_n$ can be diagonalized by a $10\times10$ unitary matrix $Z_n$:
\begin{eqnarray}
Z_n^T{M_n}{Z_n} = {M_{nd}}\:,
\end{eqnarray}
where $M_{nd}$ is the diagonal neutral fermion mass matrix. Then, we have the neutral fermion mass eigenstates:
\begin{eqnarray}
\chi _\alpha ^ \circ  = \left( {\begin{array}{*{20}{c}}
   {\kappa _\alpha ^ \circ }  \\
   { \overline{\kappa _\alpha ^ \circ} }  \\
\end{array}} \right), \quad {\alpha  = 1, \ldots, 10}
\end{eqnarray}
with
\begin{eqnarray}
\left\{ {\begin{array}{*{20}{c}}
   {{\tilde B^ \circ } = Z_n^{1\alpha }\kappa _\alpha ^ \circ \:,\quad\: {\tilde H_d} = Z_n^{3\alpha }\kappa _\alpha ^ \circ \:,\quad{\nu_{R_i}} = Z_n^{\left( {4 + i} \right)\alpha }\kappa _\alpha ^ \circ \:,\,}  \\
   {{\tilde W^ \circ } = Z_n^{2\alpha }\kappa _\alpha ^ \circ \:,\quad{\tilde H_u} = Z_n^{4\alpha }\kappa _\alpha ^ \circ \:, \quad{\nu_{L_i}} = Z_n^{\left( {7 + i} \right)\alpha }\kappa _\alpha ^ \circ \:.\:}  \\
\end{array}} \right.
\end{eqnarray}

\subsection{Charged fermion mass matrix}
Charged leptons mix with the charginos and therefore in the unrotated basis where ${\Psi ^{ - T}} = \left( { - i{{\tilde \lambda }^ - },\tilde H_d^ - ,e_{L{_i}}^ - } \right)$ and ${\Psi ^{ + T}} = \left( { - i{{\tilde \lambda }^ + },\tilde H_u^ + ,e_{R{_i}}^+} \right)$, one can obtain the charged fermion mass terms in the Lagrangian:
\begin{eqnarray}
- {\Psi ^{ - T}}{M_c}{\Psi^+} + {\rm{H.c.}}\:,
\end{eqnarray}
where
\begin{eqnarray}
{M_c} = \left( {\begin{array}{*{20}{c}}
   {{M_ \pm }} & b  \\
   c & {{m_l}}  \\
\end{array}} \right).
\end{eqnarray}
Here, the submatrix $M_ \pm $ is chargino mass matrix
\begin{eqnarray}
{M_ \pm } = \left( {\begin{array}{*{20}{c}}
   {{M_2}} & {g_2 {\upsilon_u}}  \\
   {g_2 {\upsilon_d}} & {{\lambda _i} \upsilon_{\nu_i^c}}  \\
\end{array}} \right).
\end{eqnarray}
And the submatrices $b$ and $c$ give rise to chargino-charged lepton mixing. They are defined as
\begin{eqnarray}
b = \left( {\begin{array}{*{20}{c}}
   0 & 0 & 0  \\
   { - {Y_{e_{i1}}} \upsilon_{\nu _i}} & { - {Y_{e_{i2}}} \upsilon_{\nu _i}} & { - {Y_{e_{i3}}} \upsilon_{\nu _i}}  \\
\end{array}} \right),
\end{eqnarray}
\begin{eqnarray}
c = \left( {\begin{array}{*{20}{c}}
   {g_2 \upsilon_{\nu _1}} & { - {Y_{\nu_{1i}}}\upsilon_{\nu_i^c}}  \\
   {g_2 \upsilon_{\nu _2}} & { - {Y_{\nu_{2i}}}\upsilon_{\nu_i^c}}  \\
   {g_2 \upsilon_{\nu _3}} & { - {Y_{\nu_{3i}}}\upsilon_{\nu_i^c}}  \\
\end{array}} \right).
\end{eqnarray}
And the submatrix $m_l$ is the charged lepton mass matrix
\begin{eqnarray}
{m_l} = \left( {\begin{array}{*{20}{c}}
   {{Y_{e_{11}}}{\upsilon_d}} & {{Y_{e_{12}}}{\upsilon_d}} & {{Y_{e_{13}}}{\upsilon_d}}  \\
   {{Y_{e_{21}}}{\upsilon_d}} & {{Y_{e_{22}}}{\upsilon_d}} & {{Y_{e_{23}}}{\upsilon_d}}  \\
   {{Y_{e_{31}}}{\upsilon_d}} & {{Y_{e_{32}}}{\upsilon_d}} & {{Y_{e_{33}}}{\upsilon_d}}  \\
\end{array}} \right).
\end{eqnarray}
This $5\times5$ mass matrix $M_c$ can be diagonalized by the $5\times5$ unitary matrices $Z_-$ and $Z_+$:
\begin{eqnarray}
Z_ - ^T{M_c}{Z_ + } = {M_{cd}}\:,
\end{eqnarray}
where $M_{cd}$ is the diagonal charged fermion mass matrix. Then, one can obtain the charged fermion mass eigenstates:
\begin{eqnarray}
{\chi _\alpha } = \left( {\begin{array}{*{20}{c}}
   {\kappa _\alpha ^ - }  \\
   {\overline{{\kappa _\alpha ^+}}}  \\
\end{array}} \right),\quad {\alpha  = 1, \ldots, 5}
\end{eqnarray}
with
\begin{eqnarray}
\left\{ {\begin{array}{*{20}{c}}
   {{{\tilde \lambda }^ - } = iZ_ - ^{1\alpha }\kappa _\alpha ^ - \:,\quad \tilde H_d^ -  = Z_ - ^{2\alpha }\kappa _\alpha ^ - \:,\quad {e_{L_i}} = Z_ - ^{\left( {2 + i} \right)\alpha }\kappa _\alpha ^ - ;}  \\
   \;{{{\tilde \lambda }^ + } = iZ_ + ^{1\alpha }\kappa _\alpha ^ + \:,\quad {{\tilde H}_u^+} = Z_ + ^{2\alpha }\kappa _\alpha ^ + \:,\quad {e_{R_i}} = Z_+^{\left( {2 + i} \right)\alpha }\kappa _\alpha ^+\:.\:}  \\
\end{array}} \right.
\end{eqnarray}

\section{Approximate diagonalization of mass matrices \label{appendix-approximate}}

\subsection{Neutral fermion mass matrix}
If the R-parity breaking parameters are small in the sense that for \cite{Roy,Valle2}
\begin{eqnarray}
\xi  = m.{M^{ - 1}},
\end{eqnarray}
all ${\xi _{ij}} \ll 1$, one can find an approximate diagonalization of neutral fermion mass matrix. In leading order in  $\xi$, the rotation matrix $Z_n$ is given by
\begin{eqnarray}
{Z_n} = \left( {\begin{array}{*{20}{c}}
   {1 - \frac{1}{2}{\xi ^T}\xi } & { - {\xi ^T}}  \\
   \xi  & {1 - \frac{1}{2}\xi {\xi ^T}}  \\
\end{array}} \right)\left( {\begin{array}{*{20}{c}}
   V & 0  \\
   0 & {{U_\nu }}  \\
\end{array}} \right).
\label{Zn}
\end{eqnarray}
The first matrix in (\ref{Zn}) above approximately block-diagonalizes the matrix $M_n$ to the form $diag\left( {M,{m_{eff}}} \right)$, where
\begin{eqnarray}
{m_{eff}} =  -  m.{M^{ - 1}} .{m^T}\:.
\end{eqnarray}
The submatrices $V$ and $U_{\nu}$ respectively diagonalize $M$ and ${m_{eff}}$ in the following way:
\begin{eqnarray}
\left\{ \begin{array}{l}
 {V^T}MV = {M_d}\:, \\
 U_\nu ^T{m_{eff}}{U_\nu} = {m_{\nu d}}\:, \\
 \end{array} \right.
\end{eqnarray}
where $M_d$ and ${m_{\nu d}}$ are respectively diagonal neutralino and neutrino mass matrix.

\subsection{Charged fermion mass matrix}
Similarly to the approximate diagonalization of the neutral fermion mass matrix discussed above, it's also possible to find an approximate diagonalization procedure of the charged fermion mass matrix for the small R-parity breaking parameters \cite{Valle2}. Then, we can define
\begin{eqnarray}
\left\{ \begin{array}{l}
 {\xi _L} = c.M_ \pm ^{ - 1} + {m_l}.{b^T}.{(M_ \pm ^{ - 1})^T}.M_ \pm ^{ - 1}; \\
 {\xi _R} = {b^T}.{(M_ \pm ^{ - 1})^T} + {m_l}^T.c.M_ \pm ^{ - 1}.{(M_ \pm ^{ - 1})^T}. \\
 \end{array} \right.
\end{eqnarray}
All $\xi_{L_{ij}} \ll 1$ and $\xi_{R_{ij}} \ll 1$, so in leading order in $\xi_L$ and $\xi_R$, the rotation matrices $Z_-$ and $Z_+$ are respectively given by
\begin{eqnarray}
{Z_-} = \left( {\begin{array}{*{20}{c}}
   {1 - \frac{1}{2}{\xi_L ^T}\xi_L } & { - {\xi_L ^T}}  \\
   \xi_L  & {1 - \frac{1}{2}\xi_L {\xi_L ^T}}  \\
\end{array}} \right)\left( {\begin{array}{*{20}{c}}
   U_- & 0  \\
   0 & {{V_- }}  \\
\end{array}} \right),
\end{eqnarray}
\begin{eqnarray}
{Z_+} = \left( {\begin{array}{*{20}{c}}
   {1 - \frac{1}{2}{\xi_R ^T}\xi_R } & { - {\xi_R ^T}}  \\
   \xi_R  & {1 - \frac{1}{2}\xi_R {\xi_R ^T}}  \\
\end{array}} \right)\left( {\begin{array}{*{20}{c}}
   U_+ & 0  \\
   0 & {{V_+ }}  \\
\end{array}} \right).
\end{eqnarray}
Then the matrix $M_c$ can approximately be block-diagonalized to the form $diag\left( {{M_ \pm },{m_l}} \right)$. And the submatrices $U_-,U_+$ and $V_-,V_+$ respectively diagonalize $M_\pm$ and $m_l$ in the following way:
\begin{eqnarray}
\left\{ \begin{array}{l}
 U_ - ^T{M_ \pm }{U_ + } = {M_{ \pm d}}\:, \\
 V_ - ^T{m_l}{V_ + } = {m_{ld}}\:, \\
 \end{array} \right.
\end{eqnarray}
where ${M_{ \pm d}}$ and ${m_{ld}}$ are respectively diagonal chargino and charged lepton mass matrix.

\section{Interaction Lagrangian\label{appendix-interaction}}

In this part, we give the interaction Lagrangian of the relative vertices for the LFV processes in the $\mu\nu$SSM. And we use the indices $i,j=1,\ldots,3$, $\beta,\zeta=1,\ldots,5$, $\alpha,\rho=1,\ldots,8$ and $\eta=1,\ldots,10$.

\subsection{Charged fermion-neutral fermion-gauge boson }
We now give the interaction Lagrangian of charged fermion, neutral fermion and gauge boson,
\begin{eqnarray}
&&\mathcal{L}_{int} = e F_\mu \bar{\chi}_\beta \gamma^\mu \chi_\beta + Z_\mu \bar{\chi}_\beta (C_L^{Z \chi_\zeta \bar{\chi}_\beta}\gamma^\mu P_L +  C_R^{Z \chi_\zeta \bar{\chi}_\beta}\gamma^\mu P_R) \chi_\zeta\nonumber\\
&&\qquad\quad  +\; W_\mu^+ \bar{\chi}_\eta^0 (C_L^{W \chi_\beta \bar{\chi}_\eta^0}\gamma^\mu P_L  +  C_R^{W \chi_\beta \bar{\chi}_\eta^0}\gamma^\mu P_R) \chi_\beta \nonumber\\
&&\qquad\quad  +\; W_\mu^- \bar{\chi}_\beta (C_L^{W \chi_\eta^0 \bar{\chi}_\beta}\gamma^\mu P_L +  C_R^{W \chi_\eta^0 \bar{\chi}_\beta}\gamma^\mu P_R) \chi_\eta^0 + \cdots,
\end{eqnarray}
where the coefficients are
\begin{eqnarray}
&&C_L^{Z{\chi _\zeta }{{\bar \chi }_{^\beta }}} = \frac{e}{{2{s_{_W}}{c_{_W}}}}\Big[ {( {1 - 2s_{_W}^2} ){\delta ^{\zeta \beta }} + Z{{_ - ^{1\zeta }}^ * }Z_ - ^{1\beta }} \Big]\:,  \nonumber\\
&&C_R^{Z{\chi _\zeta }{{\bar \chi }_{^\beta }}} = \frac{e}{{2{s_{_W}}{c_{_W}}}}\Big[ {2Z{{_ + ^{1\zeta }}^ * }Z_ + ^{1\beta } + Z{{_ + ^{2\zeta }}^ * }Z_ + ^{2\beta } - 2s_W^2{\delta ^{\zeta \beta }}} \Big]\:, \nonumber\\
&&C_L^{W{\chi _{^\beta }}\bar \chi _\eta ^ \circ } =  - \frac{e}{{\sqrt 2 {s_{_W}}}}\Big[ \sqrt 2 Z_ - ^{1\beta }Z{{_n^{2\eta }}^ * } + Z_ - ^{2\beta }Z{{_n^{3\eta }}^ * } + Z_ - ^{(2 + i)\beta }Z{{_n^{(7 + i)\eta }}^ * } \Big]\:,\nonumber\\
&&C_R^{W{\chi _{^\beta }}\bar \chi _\eta ^ \circ } =  - \frac{e}{{\sqrt 2 {s_{_W}}}}\Big[ \sqrt 2 Z{{_ + ^{1\beta }}^ * }Z_n^{2\eta } - Z{{_ + ^{2\beta }}^ * }Z_n^{4\eta } \Big]\:,\nonumber\\
&&C_L^{W\chi _\eta ^ \circ {{\bar \chi }_{^\beta }}} = \Big[ {C_L^{W{\chi _{^\beta }}\bar \chi _\eta ^ \circ }} \Big]^ * ,\qquad C_R^{W\chi _\eta ^ \circ {{\bar \chi }_{^\beta }}} = \Big[ {C_R^{W{\chi _{^\beta }}\bar \chi _\eta ^ \circ }} \Big]^ * .
\end{eqnarray}

\subsection{Charged scalars-gauge boson}
The interaction Lagrangian of charged scalars and gauge boson is written as
\begin{eqnarray}
\mathcal{L}_{int} = i e F_\mu S_\alpha^{-\ast}{\mathord{\buildrel{\lower3pt\hbox{$\scriptscriptstyle\leftrightarrow$}}
\over {\partial^\mu} } } S_\alpha^- + i e C^{Z S_\alpha^- S_\rho^{-\ast}} Z_\mu S_\rho^{-\ast}{\mathord{\buildrel{\lower3pt\hbox{$\scriptscriptstyle\leftrightarrow$}}
\over {\partial^\mu} } } S_\alpha^-  + \cdots.
\end{eqnarray}
The coefficient is
\begin{eqnarray}
{C^{ZS_\alpha ^ -  S_\rho ^{ -  * }}} = \frac{e}{{2{s_{_W}}{c_{_W}}}}\Big[ ( {1 - 2s_{_W}^2} ){\delta ^{\alpha \rho }} - R{{_{{S^ \pm }}^{(5 + i)\alpha }}^ * }R_{{S^ \pm }}^{(5 + i)\rho } \Big]\:.
\end{eqnarray}

\subsection{Charged fermion-neutral fermion-scalars}
The interaction Lagrangian of charged fermion, neutral fermion and scalars is similarly written by
\begin{eqnarray}
&&\mathcal{L}_{int} = S_\alpha \bar{\chi}_\zeta (C_L^{{S_\alpha }{\chi _\beta }{{\bar \chi }_\zeta }}{P_L} + C_R^{{S_\alpha }{\chi _\beta }{{\bar \chi }_\zeta }}{P_R}) \chi_\beta + P_\alpha \bar{\chi}_\zeta (C_L^{{P_\alpha }{\chi _\beta }{{\bar \chi }_\zeta }}{P_L}  \nonumber\\
&& \qquad\quad + C_R^{{P_\alpha }{\chi _\beta }{{\bar \chi }_\zeta }}P_R ) \chi_\beta + S_\alpha^- \bar{\chi}_\beta (C_L^{S_\alpha ^ - \chi _\eta ^ \circ {{\bar \chi }_\beta }}{P_L} + C_R^{S_\alpha ^ - \chi _\eta ^ \circ {{\bar \chi }_\beta }}{P_R} ) \chi_\eta^0  \nonumber\\
&& \qquad\quad + S_\alpha^{-\ast} \bar{\chi}_\eta^0 (C_L^{S_\alpha ^{-\ast} {\chi _\beta }\bar \chi _\eta ^ \circ }{P_L} + C_R^{S_\alpha ^{-\ast} {\chi _\beta }\bar \chi _\eta ^ \circ }{P_R} ) \chi_\beta  + \cdots.
\end{eqnarray}
And the coefficients are
\begin{eqnarray}
&&C_L^{{S_\alpha }{\chi _\beta }{{\bar \chi }_\zeta }} =   \frac{-e}{{{\sqrt{2}s_{_W}}}}\Big[ R_S^{2\alpha }Z_ - ^{1\beta }Z_ + ^{2\zeta } + R_S^{1\alpha }Z_ - ^{2\beta }Z_ + ^{1\zeta } + R_S^{(5 + i)\alpha }Z_ - ^{(2 + i)\beta }Z_ + ^{1\zeta } \Big]  \nonumber\\
&&\qquad\qquad\;\; + \,\frac{1}{\sqrt{2}} {Y_{e_{ij}}}\Big[ R_S^{(5 + i)\alpha }Z_ - ^{1\beta }Z_ + ^{(2 + j)\zeta } - R_S^{1\alpha }Z_ - ^{(2 + i)\beta }Z_ + ^{(2 + j)\zeta }  \Big] \nonumber\\
&&\qquad\qquad\;\; - \,\frac{1}{\sqrt{2}}{Y_{\nu_{ij}}}R_S^{(2 + j)\alpha }Z_ - ^{(2 + i)\beta }Z_ + ^{2\zeta } - \frac{1}{\sqrt{2}}{\lambda _i}R_S^{(2 + i)\alpha }Z_ - ^{2\beta }Z_ + ^{2\zeta }  \:, \nonumber\\
&&C_L^{{P_\alpha }{\chi _\beta }{{\bar \chi }_\zeta }} = \frac{{ie}}{{{\sqrt{2}s_{_W}}}}\Big[R_P^{2\alpha }Z_ - ^{1\beta }Z_ + ^{2\zeta } + R_P^{1\alpha }Z_ - ^{2\beta }Z_ + ^{1\zeta } + R_P^{(5 + i)\alpha }Z_ - ^{(2 + i)\beta }Z_ + ^{1\zeta }\Big]  \nonumber\\
&&\qquad\qquad\;\;  + \, \frac{i}{\sqrt{2}}{Y_{{e_{ij}}}}\Big[ R_P^{(5 + i)\alpha }Z_ - ^{1\beta }Z_ + ^{(2 + j)\zeta } - R_P^{1\alpha }Z_ - ^{(2 + i)\beta }Z_ + ^{(2 + j)\zeta } \Big] \nonumber\\
&&\qquad\qquad\;\;  - \, \frac{i}{\sqrt{2}}{Y_{{\nu _{ij}}}}R_P^{(2 + j)\alpha }Z_ - ^{(2 + i)\beta }Z_ + ^{2\zeta } - \frac{i}{\sqrt{2}}{\lambda _i}R_P^{(2 + i)\alpha }Z_ - ^{2\beta }Z_ + ^{2\zeta } \:, \nonumber\\
&&C_L^{S_\alpha^- \chi _\eta^0 {{\bar{\chi}}_\beta }} =   \frac{-e}{{\sqrt{2} {s_W}{c_W}}}R{_{{S^\pm }}^{2\alpha \ast } }Z_+^{2\beta} \Big[ {{c_W}Z_n^{2\eta } + {s_W}Z_n^{1\eta }} \Big]  - \frac{e}{{{s_W}}}R{_{{S^ \pm }}^{2\alpha\ast } }Z_ + ^{1\beta }Z_n^{4\eta }  \nonumber\\
&&\qquad\qquad\;\; - \frac{{\sqrt{2} e}}{{{s_W}}}R{_{{S^\pm }}^{(5 + i)\alpha\ast } }Z_ + ^{(2 + i)\beta }Z_n^{1\eta } + {Y_{\nu_{ij}}}R_{{S^ \pm }}^{(2 + i)\alpha }Z_ + ^{2\beta }Z_n^{(4 + j)\eta } \nonumber\\
&&\qquad\qquad\;\;  + \, {Y_{e_{ij}}}Z_ + ^{(2 + j)\beta } \Big[ R_{{S^ \pm }}^{1\alpha }Z_n^{(7 + i)\eta } - R_{{S^ \pm }}^{(2 + i)\alpha }Z_n^{3\eta } \Big] - {\lambda _i}R_{{S^ \pm }}^{1\alpha }Z_ + ^{2\beta }Z_n^{(4 + i)\eta },\nonumber\\
&&C_L^{S_\alpha ^{-\ast} {\chi _\beta }\bar \chi _\eta ^ \circ } =   \frac{e}{{\sqrt 2 {s_W}{c_W}}}\Big[ R{{_{{S^ \pm }}^{1\alpha\ast }} }Z_ - ^{2\beta } + R{{_{{S^ \pm }}^{(2 + i)\alpha }}^ * }Z_ - ^{(2 + i)\beta }\Big]\Big[ {c_W}Z_n^{2\eta } + {s_W}Z_n^{1\eta }\Big] \nonumber\\
&&\qquad\qquad\;\; - \frac{e}{{{s_W}}}Z_ - ^{1\beta }\Big[ R{{_{{S^ \pm }}^{1\alpha\ast }} }Z_n^{3\eta } + R{{_{{S^ \pm }}^{(2 + i)\alpha\ast }} }Z_n^{(7 + i)\eta }\Big] + {Y_{\nu_{ij}}}R_{{S^ \pm }}^{2\alpha }Z_ - ^{(2 + i)\beta }Z_n^{(4 + j)\eta }\nonumber\\
&&\qquad\qquad\;\; +\: {Y_{{e_{ij}}}}R_{{S^ \pm }}^{(5 + j)\alpha }\Big[ Z_ - ^{2\beta }Z_n^{(7 + i)\eta } - Z_ - ^{(2 + i)\beta }Z_n^{3\eta } \Big] - {\lambda _i}R_{{S^ \pm }}^{2\alpha }Z_ - ^{2\beta }Z_n^{(4 + i)\eta },\nonumber\\
&&C_R^{{S_\alpha }{\chi _\beta }{{\bar \chi }_\zeta }} = \Big[ {C_L^{{S_\alpha }{\chi _\zeta }{{\bar \chi }_\beta }}} \Big]^ * ,\qquad\quad  C_R^{{P_\alpha }{\chi _\beta }{{\bar \chi }_\zeta }} = \Big[ {C_L^{{P_\alpha }{\chi _\zeta }{{\bar \chi }_\beta }}} \Big]^ * ,\nonumber\\
&&C_R^{S_\alpha ^ - \chi _\eta ^ \circ {{\bar \chi }_\beta }} = \Big[ {C_L^{S_\alpha ^{-\ast} {\chi _\beta }\bar \chi _\eta ^0 }} \Big]^ * , \qquad\;\;\,  C_R^{S_\alpha ^{-\ast} {\chi _\beta }\bar \chi _\eta ^ \circ } =\Big[ {C_L^{S_\alpha ^ - \chi _\eta ^ \circ {{\bar \chi }_\beta }}} \Big]^ * .
\end{eqnarray}

\section{Loop-momentum integral \label{appendix-integral}}
Defining ${x_i} = \frac{{m_i^2}}{{m_W^2}}$, we can find the loop-momentum integral for $l_j^ -  \to l_i^ - \gamma$:
\begin{eqnarray}
&&{I_1}(\textit{x}_1 , x_2 ) = \frac{1}{{16{\pi ^2}}}\Big[ \frac{{1 + \ln {x_2}}}{{({x_2} - {x_1})}} + \frac{{{x_1}\ln {x_1}}-{{x_2}\ln {x_2}}}{{{{({x_2} - {x_1})}^2}}} \Big]\:,\\
&&{I_2}(\textit{x}_1 , x_2 ) = \frac{1}{{16{\pi ^2}}}\Big[ - \frac{{1 + \ln {x_1}}}{{({x_2} - {x_1})}} - \frac{{{x_1}\ln {x_1}}-{{x_2}\ln {x_2}}}{{{{({x_2} - {x_1})}^2}}} \Big]\:,\\
&&{I_3}(\textit{x}_1 , x_2 ) = \frac{1}{{32{\pi ^2}}}\Big[  \frac{{3 + 2\ln {x_2}}}{{({x_2} - {x_1})}} - \frac{{2{x_2} + 4{x_2}\ln {x_2}}}{{{{({x_2} - {x_1})}^2}}} -\frac{{2x_1^2\ln {x_1}}}{{{{({x_2} - {x_1})}^3}}} \nonumber\\
&&\qquad\qquad\quad\; + \: \frac{{2x_2^2\ln {x_2}}}{{{{({x_2} - {x_1})}^3}}}\Big]\:, \\
&&{I_4}(\textit{x}_1 , x_2 ) = \frac{1}{{96{\pi ^2}}} \Big[ \frac{{11 + 6\ln {x_2}}}{{({x_2} - {x_1})}}- \frac{{15{x_2} + 18{x_2}\ln {x_2}}}{{{{({x_2} - {x_1})}^2}}} + \frac{{6x_2^2 + 18x_2^2\ln {x_2}}}{{{{({x_2} - {x_1})}^3}}}  \nonumber\\
&&\qquad\qquad\quad\; + \: \frac{{6x_1^3\ln {x_1}}-{6x_2^3\ln {x_2}}}{{{{({x_2} - {x_1})}^4}}}  \Big]\:.\
\end{eqnarray}

And we also can find the loop-momentum integral for $l_j^-  \rightarrow l_i^- l_i^- l_i^+$:
\begin{eqnarray}
&&{G_1}(\textit{x}_1 , x_2 , x_3) =  \frac{1}{{16{\pi ^2}}}\Big[ \frac{{{x_1}\ln {x_1}}}{{({x_1} - {x_2})({x_1} - {x_3})}} + \frac{{{x_2}\ln {x_2}}}{{({x_2} - {x_1})({x_2} - {x_3})}} \nonumber\\
&&\qquad\qquad\qquad\quad +  \: \frac{{{x_3}\ln {x_3}}}{{({x_3} - {x_1})({x_3} - {x_2})}}\Big], \\
&&{G_2}(\textit{x}_1 , x_2 , x_3) =  \frac{1}{{16{\pi ^2}}}\Big[ -(\Delta  + 1 + \ln {x_\mu })  + \frac{{x_1^2\ln {x_1}}}{{({x_1} - {x_2})({x_1} - {x_3})}}  \nonumber\\
&&\qquad\qquad\qquad\quad  + \: \frac{{x_2^2\ln {x_2}}}{{({x_2} - {x_1})({x_2} - {x_3})}}+ \frac{{x_3^2\ln {x_3}}}{{({x_3} - {x_1})({x_3} - {x_2})}} \Big]\:.\qquad\quad
\end{eqnarray}
Here, ${x_\mu } = \frac{{{\mu ^2}}}{{m_W^2}}$. ${G_2}(\textit{x}_1 , x_2 , x_3)$ is divergence, so here we use dimensional regularization to cancel the divergent part $(\Delta  + 1 + \ln {x_\mu })$. In the numerical calculation, we will keep the remaining convergent part.
\begin{eqnarray}
&&{G_3}(\textit{x}_1 , x_2 , x_3, x_4) = \frac{1}{{16{\pi ^2}}}\Big[\frac{{{x_1}\ln {x_1}}}{{({x_1} - {x_2})({x_1} - {x_3})({x_1} - {x_4})}}  \nonumber\\
&&\qquad\qquad  + \frac{{{x_2}\ln {x_2}}}{{({x_2} - {x_1})({x_2} - {x_3})({x_2} - {x_4})}} + \frac{{{x_3}\ln {x_3}}}{{({x_3}  - {x_1})({x_3} - {x_2})({x_3} - {x_4})}} \nonumber\\
&&\qquad\qquad  + \: \frac{{{x_4}\ln {x_4}}}{{({x_4} - {x_1})({x_4} - {x_2})({x_4} - {x_3})}}\Big] \:, \\
&&{G_4}(\textit{x}_1 , x_2 , x_3, x_4) = \frac{1}{{16{\pi ^2}}}\Big[\frac{{x_1^2\ln {x_1}}}{{({x_1} - {x_2})({x_1} - {x_3})({x_1} - {x_4})}}  \nonumber\\
&&\qquad\qquad  + \frac{{x_2^2\ln {x_2}}}{{({x_2} - {x_1})({x_2} - {x_3})({x_2} - {x_4})}} + \frac{{x_3^2\ln {x_3}}}{{({x_3}  - {x_1})({x_3} - {x_2})({x_3} - {x_4})}}   \nonumber\\
&&\qquad\qquad + \: \frac{{x_4^2\ln {x_4}}}{{({x_4} - {x_1})({x_4} - {x_2})({x_4} - {x_3})}}\Big]\:.
\end{eqnarray}

\bibliographystyle{model1-num-names}

\end{document}